\documentclass[journal,twoside,web]{ieeecolor}
\usepackage{generic}
\usepackage{cite}
\usepackage{amsmath,amssymb,amsfonts}
\usepackage{algorithmic}
\usepackage{graphicx}
\usepackage{algorithm,algorithmic}
\usepackage{hyperref}
\usepackage{multirow}
\usepackage{tabularx}
\usepackage[normalem]{ulem}
\useunder{\uline}{\ul}{}
\hypersetup{hidelinks=true}
\usepackage{textcomp}
\def\BibTeX{{\rm B\kern-.05em{\sc i\kern-.025em b}\kern-.08em
    T\kern-.1667em\lower.7ex\hbox{E}\kern-.125emX}}
\begin{document}
\title{Multi-Site rs-fMRI Domain Alignment for Autism Spectrum Disorder Auxiliary Diagnosis Based on Hyperbolic Space}
\author{Yiqian Luo, Qiurong Chen, Fali Li, Peng Xu, \IEEEmembership{Member, IEEE}, Yangsong Zhang
\thanks{This work was supported in part by the National Natural Science Foundation of China under Grant {\#}62076209 and {\#}W2411084, and the Key R\&D projects of the Science \& Technology Department of Chengdu ({\#}2024-YF08-00072-GX). (Corresponding authors: Fali Li, Peng Xu, Yangsong Zhang.)}
\thanks{Yiqian Luo, Qiurong Chen are with the Laboratory for Brain Science and Artificial Intelligence, School of Computer Science and Technology, Southwest University of Science and Technology, Mianyang 621010, China.}
\thanks{Yangsong Zhang is with the Laboratory for Brain Science and Artificial Intelligence, School of Computer Science and Technology, is also with School of Medicine, Southwest University of Science and Technology, Mianyang 621010, China (e-mail:zhangysacademy@gmail.com).}
\thanks{Fali Li is with the MOE Key Laboratory for Neuroinformation, Clinical Hospital of Chengdu Brain Science Institute, University of Electronic Science and Technology of China, Chengdu 610054, China (e-mail:fali.li@uestc.edu.cn).}
\thanks{Peng Xu are with School of Computer Science and Technology, Southwest University of Science and Technology, Mianyang 621010, China, and also with the MOE Key Laboratory for Neuroinformation, Clinical Hospital of Chengdu Brain Science Institute, University of Electronic Science and Technology of China, Chengdu 610054, China (e-mail:xupeng@uestc.edu.cn).}}

\maketitle

\begin{abstract}
Increasing the volume of training data can enable the auxiliary diagnostic algorithms for Autism Spectrum Disorder (ASD) to learn more accurate and stable models. However, due to the significant heterogeneity and domain shift in rs-fMRI data across different sites, the accuracy of auxiliary diagnosis remains unsatisfactory. Moreover, there has been limited exploration of multi-source domain adaptation models on ASD recognition, and many existing models lack inherent interpretability, as they do not explicitly incorporate prior neurobiological knowledge such as the hierarchical structure of functional brain networks. To address these challenges, we proposed a domain-adaptive algorithm based on hyperbolic space embedding. Hyperbolic space is naturally suited for representing the topology of complex networks such as brain functional networks. Therefore, we embedded the brain functional network into hyperbolic space and constructed the corresponding hyperbolic space community network to effectively extract latent representations. To address the heterogeneity of data across different sites and the issue of domain shift, we introduce a constraint loss function, Hyperbolic Maximum Mean Discrepancy (HMMD), to align the marginal distributions in the hyperbolic space. Additionally, we employ class prototype alignment to mitigate discrepancies in conditional distributions across domains. Experimental results indicate that the proposed algorithm achieves superior classification performance for ASD compared to baseline models, with improved robustness to multi-site heterogeneity. Specifically, our method achieves an average accuracy improvement of 4.03\%. Moreover, its generalization capability is further validated through experiments conducted on extra Major Depressive Disorder (MDD) datasets. The code is available at https://github.com/LYQbyte/H2MSDA.
\end{abstract}


\section{Introduction}
\label{sec:introduction}
\IEEEPARstart{A}{utism} Spectrum Disorder (ASD) is a widespread developmental disorder that affects approximately 1\% of the population~\cite{pandolfi2018screening}, severely disrupting daily life and being associated with a high prevalence of comorbid conditions such as depression and anxiety~\cite{vohra2017comorbidity}. ASD typically emerges in childhood, with lifelong symptoms that often require long-term support~\cite{lord2018autism}. The main characteristics of ASD include social communication deficits and restricted, repetitive behaviors, with severity ranging from mild to profound~\cite{hirota2023autism}. Currently, there are still no effective medications for treating ASD, early and accurate diagnosis and intervention are of great importance for ASD. However, existing diagnostic methods such as questionnaires and behavioral observation are subjective and influenced by environmental factors, increasing the risk of misdiagnosis. Therefore, there is an urgent need for more accurate diagnostic methods and the identification of additional biomarkers to aid in diagnosis.

With the rapid development of medical imaging technologies, tools for analyzing neurophysiology have become increasingly important~\cite{Ye2024MAD}. Resting-state functional magnetic resonance imaging (rs-fMRI) is a non-invasive technique~\cite{matthews2004functional} with high spatial resolution. It captures intrinsic brain activity by measuring Blood Oxygen Level Dependent (BOLD) signal fluctuations induced by neuronal activity~\cite{friston1994analysis}, thereby reflecting underlying functional connectivity without task-related constraints. Many studies have demonstrated the feasibility of using rs-fMRI to explore brain network interactions in patients with psychiatric disorders~\cite{woodward2015resting,FANG2025Source}. In recent years, numerous researchers have combined deep learning techniques with rs-fMRI to explore psychiatric disorders, achieving significant results~\cite{dong2020compression, luppi2022synergistic}. Most studies use pre-defined brain templates to divide the brain into regions of interest (ROIs), obtain the BOLD time series of these ROIs, and then compute the correlations between the time series to construct functional connectivity networks (FCNs), enabling brain activity analysis~\cite{dong2024brain, wang2023covariance}.

In ASD diagnosis using rs-fMRI data, researchers often integrate datasets from multiple medical sites to address the scarcity of training data~\cite{SONG2024BrainDAS, YE2025Fuse}. However, due to the heterogeneity in equipment, scanning parameters, and data characteristics across different medical sites, there are significant differences in the data distributions between sites, leading to the domain shift problem~\cite{FANG2025Source}. This discrepancy severely limits the generalization ability of models, especially when diagnostic labels are absent for data from certain medical sites, making traditional supervised learning methods difficult to apply directly. To address this issue, Domain Adaptation (DA)~\cite{Kouw2019Review} techniques have emerged. The goal of DA is to enable knowledge transfer from the source domain to the target domain through techniques like feature alignment, thereby improving classification performance in the target domain. This approach provides a novel perspective for ASD diagnosis based on multi-site data. Currently, depending on the number of source domains, DA methods can be categorized into Single-Source Domain Adaptation (SSDA)~\cite{Zhao2020aReview} and Multi-Source Domain Adaptation (MSDA)~\cite{SUN2015survey}.

Single-source domain adaptation, involves a source domain dataset with labeled data from a single site and a target domain dataset from an unlabeled site. The goal is to reduce the feature distribution difference between the source and target domains, allowing the model to perform well in classifying data from the target domain. Chu et al.~\cite{chu2022resting} proposed an adaptive framework based on Attention Graph Convolutional Networks (A$^{2}$GCN) for two-site rs-fMRI analysis. By integrating attention mechanisms and minimizing domain discrepancies via Mean Absolute Error (MAE) and CORAL~\cite{sun2016deep}, the model enhances ASD classification performance and supports the identification of pathological brain regions. Fang et al.~\cite{FANG2023Unsupervised} developed an unsupervised cross-domain framework using attention-guided graph convolution and Maximum Mean Discrepancy (MMD)~\cite{Gretton2006Kernel} to align spatiotemporal features across domains, achieving effective depression recognition and functional connectivity localization in rs-fMRI data from 681 subjects.

SSDA methods have shown promise in brain disease diagnosis using two-site fMRI data by aligning features between a labeled source domain and an unlabeled target domain. However, SSDA is limited to simple two-site scenarios and fails to leverage the rich and diverse information from multiple source domains, which is essential in real-world settings where data are collected from various sites with significant heterogeneity. To address this, multi-source domain adaptation (MSDA) methods have been proposed to integrate knowledge across multiple labeled source domains and enhance generalization to an unlabeled target domain through more robust feature alignment.

For instance, Li et al.~\cite{LI2020Multisite} proposed a privacy-preserving federated learning framework combined with domain adaptation techniques for multi-site fMRI data analysis, specifically for ASD diagnosis. This method trains models locally at each site and aggregates updates into a global model, thus protecting patient privacy. It incorporates domain adversarial techniques~\cite{Ganin2016Domain} and mixture of experts~\cite{masoudnia2014mixture} to reduce the distribution differences across sites, providing a new perspective for ASD diagnosis with multi-source data. Other representative approaches, such as BrainDAS\cite{SONG2024BrainDAS} and LRCDR~\cite{Liu2023DomainA}, explore structure-aware alignment and low-rank discriminative representations, respectively, to better capture brain network characteristics and reduce domain shifts. These studies collectively demonstrate the potential of MSDA in enhancing classification accuracy and generalization ability in multi-site brain disease diagnosis, especially for complex neuropsychiatric disorders such as ASD.

Deep learning algorithms based on domain adaptation have demonstrated significant potential in the analysis of multi-site fMRI data. These methods effectively address issues such as discrepancies in data distribution across sites and the absence of labeled data in the target domain, thereby greatly enhancing model adaptability and classification performance in the target domain. However, most existing approaches operate in Euclidean space, which may not be optimal for representing the complex and hierarchical topology inherent in brain functional networks. Functional connectivity networks often exhibit latent hierarchical or tree-like structures, which are poorly preserved under Euclidean embeddings~\cite{Baker2024Hyperbolic}. In contrast, hyperbolic space provides a more suitable geometric foundation for modeling such data, due to its exponential growth property and ability to capture hierarchical relationships with lower distortion~\cite{Dai2021CVPR}. Leveraging hyperbolic embeddings can therefore lead to more compact and informative representations, improving the model’s ability to align multi-site distributions and uncover clinically relevant patterns.


To address the limitations of existing methods, which primarily operate in Euclidean space and thus exhibit poor representational performance when modeling brain functional connectivity networks and insufficient attention to multi-source domain adaptation. We propose a domain adaptation algorithm based on hyperbolic space embedding. The hyperbolic space naturally captures the topological structure of brain functional networks. We leverage this property to embed brain functional networks into hyperbolic space, constructing community networks in the hyperbolic domain for effective brain function state representation. To tackle the heterogeneity of multi-site data, we propose a Hyperbolic Maximum Mean Discrepancy (HMMD) for marginal distribution alignment (global alignment) between source and target domains. For conditional distribution alignment, we align the class prototypes derived from source domain labels with those generated from pseudo-labels in the target domain, ensuring more precise feature alignment.

Our main contributions are summarized as follows:
\begin{enumerate}

\item We propose H$^2$MSDA (Hyperbolic Hierarchical Multi-Sites Domain Alignment), a novel framework that integrates hyperbolic embeddings with hierarchical domain alignment to model brain functional networks. By leveraging the geometric advantages of hyperbolic space, H$^2$MSDA effectively captures complex topological structures and addresses inter-site variability in multi-site fMRI data.


\item We design a Hyperbolic Maximum Mean Discrepancy (HMMD) module to reduce global marginal distribution discrepancies across sites. In addition, we introduce a category prototype alignment strategy that aligns source domain prototypes with pseudo-label prototypes from the target domain. Together, these components enable both marginal and conditional distribution alignment, significantly enhancing cross-site generalization and classification performance.



\item Extensive experiments on the ABIDE-I and Rest-meta-MDD datasets validate the superiority and generalizability of the proposed approach, confirming its ability to effectively align domain distributions across multiple acquisition sites.

\end{enumerate}

\section{Preliminaries}
This section provides a concise overview of the key prior knowledge underpinning our research. We begin with an introduction to hyperbolic embeddings and hyperbolic graph convolutional networks, followed by a discussion on functional gradients and their application in relevant domains.

\subsection{Hyperbolic Embedding and Hyperbolic Graph Convolutional Networks}
This subsection introduces hyperbolic embeddings, focusing on the Poincar$\acute{e}$ disk model, and their integration into Graph Convolutional Networks (GCNs), highlighting how hyperbolic geometry facilitates effective representation of complex network structures.

\subsubsection{Hyperbolic Embedding}

In many real-world networks, such as social, communication, and biological systems relationships between nodes often exhibit hierarchical or non-Euclidean structures~\cite{yang2024hypformer, gu2024Unsupervised, yang2023Hyperbolic}. Traditional Euclidean embeddings struggle to capture such patterns due to the curse of dimensionality. In contrast, hyperbolic geometry, particularly the Poincar$\acute{e}$ disk model~\cite{Nickel2017Poincar}, offers a compact and effective representation for hierarchical data, as distances in hyperbolic space grow exponentially~\cite{Dai2021CVPR}.

\begin{figure}[!htbp]
	\centering
	\includegraphics[width=\linewidth]{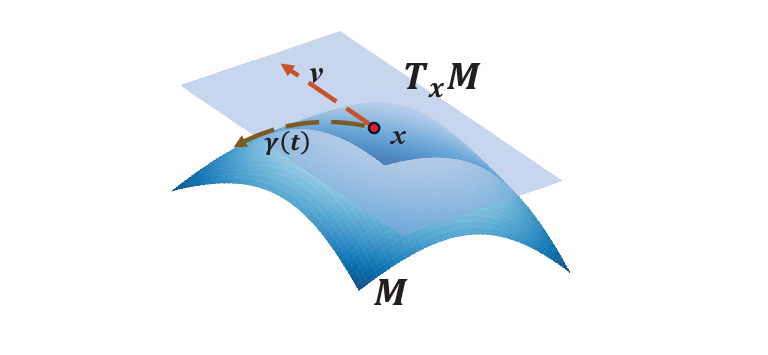}
	\caption{The tangent space $T_xM$ is a vector space associated with a specific point $x$ on a curve that lies within the manifold $M$. A tangent vector $v$ at $x$ resides in the tangent space $T_xM$ and represents the direction of the curve as it passes through $x$.}
	\label{fig1}
\end{figure}

The Poincar$\acute{e}$ disk model defines a hyperbolic space with constant negative curvature, where the set of points is confined within a unit disk~\cite{yang2023Hyperbolic}. Denote $\kappa$ as the curvature, following previous works~\cite{Chami2019Hyperbolic, yang2023Hyperbolic}, we set $\kappa=-1$ to define a fixed hyperbolic geometry in the Poincar$\acute{e}$ ball model. And ${\left\| . \right\|_2}$ as the standard euclidean inner product, for the unit disk ${D_{\kappa}} = \left\{ {x \in {R^n}:{{\left\| x \right\|}_2} < \frac{1}{{\sqrt {\kappa} }}} \right\}$, the metric defines the distance between any two points $u$ and $v$ as formula (\ref{formula1}):

\begin{equation}
	d\left( {u,v} \right) = \frac{2}{{\sqrt {\kappa} }}ar\tanh \left( {\sqrt {\kappa} {{\left\| { - u \oplus v} \right\|}_2}} \right)
	\label{formula1}
\end{equation}
where $\oplus$ stands for the Möbius addition:

\begin{equation}
	u \oplus v = \frac{{\left( {1 + 2{\kappa}\left\langle {u,v} \right\rangle  + {\kappa}\left\| v \right\|_2^2} \right)u + \left( {1 - {\kappa}\left\| u \right\|_2^2} \right)v}}{{1 + 2{\kappa}\left\langle {u,v} \right\rangle  + {{\kappa}^2}\left\| u \right\|_2^2\left\| v \right\|_2^2}}
	\label{formula2}
\end{equation}

For feature vectors or time series in Euclidean space, embedding in the Poincar$\acute{e}$ disk can be achieved through the corresponding mapping function as formula (\ref{formula3}). Conversely, embedding from the Poincar$\acute{e}$ disk to Euclidean space can be accomplished through the mapping function as formula (\ref{formula4}). These bidirectional mappings between Euclidean and hyperbolic spaces are visualized in Figure \ref{fig1}. This approach follows previous literature~\cite{Ganea2018Hyperbolic}, and ensures that feature structure is preserved during transformation, as the exponential and logarithmic maps on a hyperbolic manifold are mathematically well-defined and designed to minimize information distortion.

\begin{equation}
	\exp _0^{\kappa}\left( x \right) = \tanh \left( {\sqrt {\kappa} {{\left\| x \right\|}_2}} \right)\frac{x}{{\sqrt {\kappa} \left\| x \right\|}}
	\label{formula3}
\end{equation}

\begin{equation}
	\log _0^{\kappa}\left( x \right) = \frac{1}{{\sqrt {\kappa} }}ar\tanh \left( {\sqrt {\kappa} {{\left\| x \right\|}_2}} \right)\frac{x}{{{{\left\| x \right\|}_2}}}
	\label{formula4}
\end{equation}

\begin{table*}[]
	\renewcommand{\arraystretch}{1.3}
	\caption{Demographic Information of the Four Sites Involved in the ABIDE-I Dataset.}
	\centering
	\begin{tabular}{llccccccclc}
		\hline
		\multirow{2}{*}{Site} &  & \multicolumn{3}{c}{ASD}        &  & \multicolumn{3}{c}{HC}         &  & \multirow{2}{*}{Scanner} \\ \cline{3-5} \cline{7-9}
		&  & Age (Mean±std) &  & Gender (M/F) &  & Age (Mean±std) &  & Gender (M/F) &  &                          \\ \hline
		NYU                   &  & 14.92±7.04    &  & 64/9        &  & 15.67±6.19    &  & 72/26       &  & SIEMENS Allegra          \\
		UM                    &  & 13.85±2.29    &  & 39/9        &  & 15.03±3.64    &  & 49/16       &  & GE Signa                 \\
		USM                   &  & 24.60±8.46    &  & 38/0        &  & 22.33±7.70    &  & 23/0        &  & SIEMENS Trio             \\
		UCLA                  &  & 13.34±2.56    &  & 34/2        &  & 13.18±1.76    &  & 33/6        &  & SIEMENS Trio             \\ \hline
	\end{tabular}
	\label{table1}
\end{table*}

\begin{table}[]
	\caption{Notation table.}
	\renewcommand{\arraystretch}{1.2}
	\resizebox{\columnwidth}{!}{
	\begin{tabular}{cc}
		\hline
		\textbf{Notation}                                                              & \textbf{Description}                                                            \\ \hline
		$d(\cdot, \cdot)$                                                              & Hyperbolic distance function.                                                   \\
		$\kappa$                                                                       & The curvature of hyperbolic space.                                              \\
		$\oplus$, $\otimes$                                                            & Möbius addition and Möbius Scalar Multiplication.                               \\
		$\exp _0^{\kappa}\left( \cdot \right)$, $\log _0^{\kappa}\left( \cdot \right)$ & Mappings between hyperbolic space and Euclidean space.                \\
		$Proj$                                                                         & Projection keeping the result within the Poincar$\acute{e}$ disk. \\
		$\sigma$                                                                       & Activation function.                                                            \\
		$HLinear$                                                                      & Hyperbolic linear layer.                                                        \\
		$V$, $V_R$, $V_C$                                                              & Node features.                                                                  \\
		$A_R$, $A_C$, $A_{C-base}$                                                     & Adjacency matrix.                                                           \\
		$D$                                                                            & Degree matrix.                                                                  \\
		$W_R$, $W_C$, ${W^l}$, $b^l$                                                   & Parameter matrix.                                                               \\
		$u, v$                                                                         & Nodes in hyperbolic space.                                                      \\
		$C_i$                                                                          & $i$-th community representation.                                                \\
		$cov(\cdot ,\cdot)$                                                            & Covariance between the rank vectors                                             \\
		$\gamma_i$                                                                     & Bandwidth for $i$-th kernel.                                                    \\
		$S$, $T$                                                                       & Source domain, target domain                                                    \\
		$ROI{s_i}$                                                                     & Feature matrix of the ROIs belonging to the $i$-th community                    \\
		$M_C$                                                                          & Attention matrix                                                                \\
		$H_R$, $H_C$                                                                   & Node representation in ROI level and community level.                           \\
		$c$,  $N_c$                                                                    & The class index, number of samples in class $c$.                                \\
		$N_{class}$                                                                    & Total number of classes.                                                        \\
		$Pt_c^{source}$, $Pt_c^{target}$                                               & $c$-th class prototype of source/target.                                        \\
		$Pt_c$,                                                                        & Prototype of class $c$.                                                         \\
		$N_{gra}$                                                                      & Number of top components of functional gradient.                                \\
		$k$                                                                            & Number of Communities.                                                          \\
		$\lambda _m$, $\lambda _{PA}$                                                  & Weight hyperparameters.                                                         \\
		$N$                                                                            & Total number of samples                                                         \\
		$y_i$, $p_i$                                                                   & Ground-truth label and predicted label of the $i$-th sample.                    \\ 
		$E$																			   & Mathematical expectation.														\\
		\hline
	\end{tabular}}
\end{table}

\subsubsection{Hyperbolic Graph Convolutional Networks}
Graph Convolutional Networks are powerful deep learning methods for handling graph data. However, traditional GCNs are typically based on Euclidean space, which fails to fully leverage the hierarchical structure and complex relationships inherent in graph data. Extending GCNs to hyperbolic space allows for more efficient capture of the hierarchical relationships within the graph~\cite{Baker2024Hyperbolic}.

In the process of transforming features in hyperbolic space, the traditional matrix multiplication used in Euclidean space is no longer applicable. Instead, the feature transformation in hyperbolic space is defined as follows:
\begin{equation}
	{x^l} = Proj\left( {{W^l} \otimes {x^{l - 1}} \oplus {b^l}} \right)
	\label{formula5}
\end{equation}
where $Proj$ denotes the projection operation, which ensures that the result remains within the Poincar$\acute{e}$ disk. $W^l$ denotes the weight matrix and $b^l$ represents the bias vector. The symbol $\otimes$ represents the Möbius Scalar Multiplication as shown in formula (\ref{formula6}).

\begin{equation}
	\resizebox{.85\hsize}{!}{$
	W \otimes x = \tanh \left( {\frac{{{{\left\| {x{W^T}} \right\|}_2}}}{{{{\left\| x \right\|}_2}}}ar\tanh \left( {\sqrt {\kappa} {{\left\| x \right\|}_2}} \right)} \right)\frac{{x{W^T}}}{{\sqrt {\kappa} {{\left\| {x{W^T}} \right\|}_2}}}$}
	\label{formula6}
\end{equation}

For a graph $G(V, A)$ ($V$ represents the node features and $A$ represents the adjacency matrix of the graph), a single layer of graph convolution on the hyperbolic manifold can be represented as:

\begin{equation}
	H = \exp _0^{\kappa}\left( {\sigma \left( {A\left( {\log _0^{\kappa}\left( {HLinear\left( {V,W} \right)} \right)} \right)} \right)} \right)
	\label{formula7}
\end{equation}

Here, we use $HLinear$ to represent the feature transformation process (as formula (\ref{formula5})) in hyperbolic space, where $\sigma$ denotes the corresponding activation function.

\subsection{Brain Functional Gradients}
The brain's functional gradients, by capturing the continuous spatial patterns of extrinsic connectivity between isolated networks, offers a more comprehensive perspective of brain organization~\cite{guell2018functional, huntenburg2018large}. In the gradient-based approach, a nonlinear decomposition of high-dimensional resting-state functional connectivity, can identify the functional hierarchy of the brain by representing its connections in a continuous low-dimensional space~\cite{dong2020compression}. Functional gradients have been widely applied in the context of various psychiatric disorders, and can reveal abnormal functional hierarchies across brain regions~\cite{dong2020compression,guo2023functional,gong2023connectivity}.

Functional gradients have also been used in the study of ASD. For instance, Hong et al.~\cite{hong2019atypical} identified functional gradient disruptions in individuals with autism, indicating a reduction in functional independence between regions, particularly between unimodal and transmodal regions. Urchs et al.~\cite{urchs2022functional} further explored functional gradient analysis in ASD using the ABIDE-I dataset~\cite{di2014autism}, revealing gradient compression, which suggests reduced functional segregation in ASD patients. Building on our previous research~\cite{luo2024hierarchical}, we apply functional gradients in the same manner to partition brain community networks. This approach leverages the functional relationships between ROIs to construct corresponding functional community networks.

In previous methods for calculating functional gradients, BrainSpace~\cite{vos2020brainspace} was typically used for direct computation. However, since the functional connectivity network in this study is embedded in hyperbolic space, while traditional methods are primarily designed for Euclidean space, adjustments were necessary in our approach. Specifically, for the computation of the affinity matrix ($Af{f_{spearman}}$), we utilized the Spearman correlation coefficient (as shown in formula (\ref{aff_spear})), rather than the cosine similarity, which is commonly used in Euclidean space but is not suitable for hyperbolic geometry. This modification allows us to more accurately capture the non-Euclidean relationships between brain regions within hyperbolic space.

\begin{equation}
	Af{f_{spearman}}\left( {i,j} \right) = \frac{{{\mathop{\rm cov}} \left( {ri,rj} \right)}}{{{\sigma _{ri}}{\sigma _{rj}}}}
	\label{aff_spear}
\end{equation}
where $ri$ and $rj$ represent the rank vectors of variables $i$ and $j$, respectively, $cov$ denotes the covariance between the rank vectors, and $\sigma _{ri}$ and $\sigma _{rj}$ indicates the standard deviation of the rank vectors.

Before performing the nonlinear dimensionality reduction on the affinity matrix $Af{f_{spearman}}$, the sparsification operation was first applied. Specifically,  only the top 10\%~\cite{vos2020brainspace,gong2023connectivity} of the strongest functional connections for each ROI were kept, thereby extracting the most significant connections. Then, we normalized the sparsified affinity matrix using following formula:
\begin{equation}
	W\left( a \right) = {D^{ - \frac{1}{\alpha }}} \cdot Af{f_{spearman}} \cdot {D^{ - \frac{1}{\alpha }}}
	\label{dm_1}
\end{equation}
where $D$ denotes the degree matrix of the affinity matrix, and $\alpha$ represents the diffusion parameter, which is empirically set to a common value of 0.5~\cite{vos2020brainspace}.

Subsequently, we use diffusion mapping to extract the components of the functional gradient. This approach enables the characterization of nonlinear relationships inherent in the data, where $D_{\alpha}$ represents the diagonal matrix, with ${D_{\alpha (i,i)}} = \sum\limits_{j = 1}^{{N_R}} {{W_{i,j}}}$.

\begin{equation}
	P\left( \alpha  \right) = {D_\alpha^{ - 1}}W\left( \alpha  \right)
	\label{dm_2}
\end{equation}

Then, we perform eigenvalue decomposition to extract the top $n$ largest eigenvalues $\lambda$ and their corresponding eigenvectors $v$.
\begin{equation}
	P\left( \alpha  \right)v = \lambda v
	\label{dm_3}
\end{equation}

Finally, the diffusion embedding is constructed by scaling the eigenvectors to obtain the $i$-th functional gradient component ${\Phi _i}$.
\begin{equation}
	{\Phi _i} = \frac{{{\lambda _i}}}{{1 - {\lambda _i}}} \cdot {v_i}
	\label{dm_4}
\end{equation}


\begin{figure*}[!htbp]
	\centering
	\includegraphics[width=\linewidth]{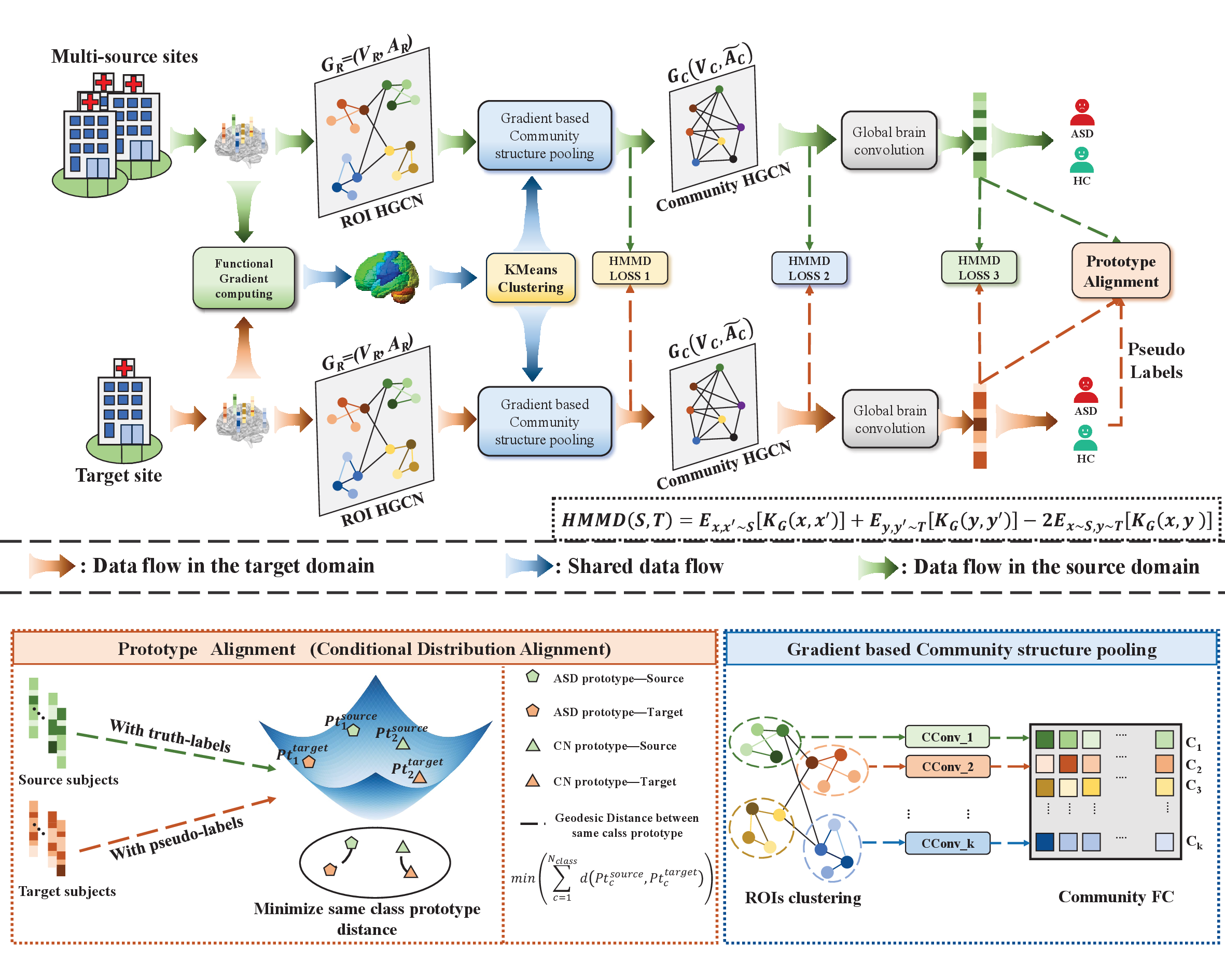}
	\caption{The diagram of the proposed Hyperbolic Hierarchical Multi-Sites Domain Alignment (H$^2$MSDA) method for ASD identification.}
	\label{model_fig}
\end{figure*}

\section{Materials and Proposed Method}
\subsection{Materials and data preprocessing}
In this study, the data from the four largest sites in the ABIDE (Autism Brain Imaging Data Exchange) dataset~\cite{di2014autism}, i.e., NYU, UM, UCLA and USM, were selected based on their substantial sample sizes, comprising 195 individuals with ASD and 225 healthy controls (HCs). These sites offer a robust foundation for evaluating domain adaptation performance. Additionally, the significant heterogeneity of the data across different sites creates an ideal experimental setting to validate the effectiveness of the domain adaptation methods. The statistical information for the data selected from each site is shown in Table \ref{table1}.

\begin{figure}[!htbp]
	\centering
	\includegraphics[width=\linewidth]{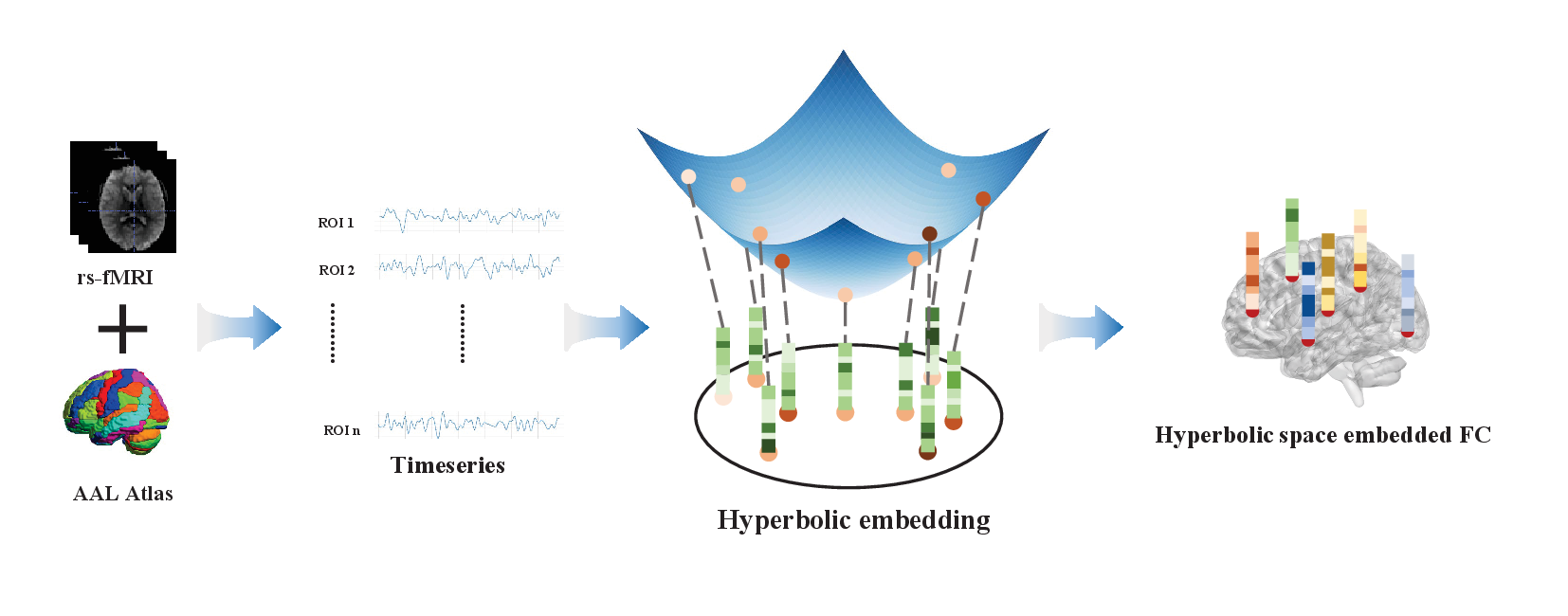}
	\caption{The pipelines process and computation of functional connectivity network.}
	\label{Figure1}
\end{figure}

We used the resting-state fMRI data processing toolbox (DPARSF) for standardization and preprocessing of the rs-fMRI data~\cite{yan2010dparsf}. The specific preprocessing steps are as follows:
\begin{enumerate}
\item Discarding the first 5 time points to ensure signal stability. 
\item Performing slice timing correction to correct for the acquisition delays between slices.
\item Applying motion correction to reduce the influence of head movement on the signals.
\item Normalizing the images to the MNI space using the EPI template and resample them to a $3\times3\times3$ mm resolution.
\item Applying spatial smoothing using a Gaussian kernel with a full-width at half-maximum (FWHM) of 4 mm.
\item Performing linear detrending and apply band-pass filtering (0.01–0.10 Hz) to extract the relevant BOLD signals.
\item Removing nuisance signals, including head motion parameters, white matter signals, cerebrospinal fluid (CSF) signals, and global signals.
\end{enumerate}
After completing the above preprocessing steps, the normalized fMRI data were partitioned into 116 ROIs using the AAL atlas~\cite{tzourio2002automated}, and the BOLD time series of each ROI were extracted. To avoid amplitude differences between different ROIs affecting subsequent analysis, the BOLD time series of each ROI were standardized, enhancing the model's robustness and the effectiveness of feature extraction.

\subsection{Proposed Method}
In the proposed method, the brain functional networks were embedded into hyperbolic space, owing to its efficient representation capability. This operation can achieve precise representation of brain functional networks, and enable effective transfer between the source domain and target domain during domain adaptation procedure. Additionally, this method significantly alleviates the data heterogeneity problem across different sites, improving the model's generalization ability and diagnostic accuracy. The proposed method is shown in Figure \ref{model_fig}.

Specifically, for the feature extraction of brain functional networks, we used Hyperbolic space based Graph Convolutional Networks (HGCN) to enable effective message passing at the ROIs level. Then, we applied functional gradients (which are derived from the hyperbolic embedding and computed in hyperbolic space) to explore and construct community networks. Through the construction of these community networks, we obtained community-level functional networks. Similarly, we employed HGCN to facilitate message passing between the communities. Finally, through whole-brain scale convolutions, we extracted vector-level representations of the entire brain network, enabling the final diagnostic classification. To address the heterogeneity issue in multi-site data, we introduce HMMD to align the marginal distributions across sites. Furthermore, prototype alignment is employed to achieve conditional distribution alignment, thereby effectively mitigating domain shift and heterogeneity in multi-site datasets.

\subsubsection{Hyperbolic Embedding and ROI-Level Feature Extraction}
After preprocessing, we obtained the BOLD time series for each ROI. To effectively construct the hyperbolic embedded functional network, we first embedded the BOLD time series of each brain region into the Poincar$\acute{e}$ disk. Then, in order to build an effective functional network, as in~\cite{rodriguez2020hyperbolic}, we selected the Spearman correlation coefficient (as formula (\ref{formula8})) to compute the pairwise correlation between the time series of brain regions. This choice was made instead of using the Pearson correlation coefficient, which is typically used for calculating functional connectivity in Euclidean space. Pearson correlation measures linear relationships, which are not suitable for manifold spaces. In contrast, the Spearman correlation can capture non-linear relationships, making it more appropriate for our manifold-based approach.

\begin{equation}
	\rho  = \frac{{{\mathop{\rm cov}} \left( {rx,ry} \right)}}{{{\sigma _{rx}}{\sigma _{ry}}}}
	\label{formula8}
\end{equation}
where $rx$ and $ry$ represent the rank vectors of variables $x$ and $y$, respectively, $cov(\cdot ,\cdot)$ denotes the covariance between the rank vectors, and $\sigma _{rx}$ and $\sigma _{ry}$ indicates the standard deviation of the rank vectors.

Through the steps outlined above, we obtained the FC embedded in the Poincar$\acute{e}$ disk for each subject. To enable effective message passing between the ROIs, we employed the Hyperbolic Graph Convolutional Network on the hyperbolic space.

Considering the presence of spurious connections in the functional connectivity network, following~\cite{Tong2023fMRI}, we empirically retain the top 10\% of the strongest connections in original FC matrix to obtain the adjacency matrix $A_R$. The hyperbolic embedded FC serves as the node features $V_R$ in the graph network. This results in a corresponding ROI-level graph ${G_R}\left( {{V_R},{A_R}} \right)$. To enable message passing between the ROIs, we applied a single layer of Hyperbolic Graph Convolutional Network, as shown in the following formula (\ref{formula9}):

\begin{equation}
	H_R = \exp _0^{\kappa}\left( {\sigma \left( {A_R\left( {\log _0^{\kappa}\left( {HLinear\left( {V_R,W_R} \right)} \right)} \right)} \right)} \right)
	\label{formula9}
\end{equation}

Functional connectivity during resting state exhibits a significant modular structure, which facilitates efficient information communication and cognitive functions~\cite{LIAO2017Individual,sporns2016modular}. Community networks play a critical role in understanding the functional organization of the brain~\cite{van2009functionally}. Functional communities have been shown to be closely related to cognitive behaviors~\cite{van2010exploring}, mental states~\cite{geerligs2015state}, and neuropsychiatric disorders~\cite{canario2021review}. Moreover, recent studies have confirmed that the brain's modularity can serve as a unified biomarker for intervention-related plasticity~\cite{gallen2019brain,Wang2024Leveraging}.

Generally, each community is composed of tightly connected brain ROIs, which are sparsely connected to ROIs in other communities~\cite{sporns2016modular}. The functional similarity between these ROIs can be effectively captured using the functional gradient. Similar to our previous work~\cite{luo2024hierarchical}, we input the group-level average FC into BrainSpace to obtain functional gradients for multiple components. The gradients of the top $N_{gra}$ components are then used for KMeans clustering, enabling the assignment of the 116 ROIs to $k$ communities.

To achieve effective feature representation of the community network, we proposed the $CConv$ module, which is specific to each community. Specifically, we used a conv1D to extract features from the ROIs that belong to the same community, thereby achieving the feature representation of the community, as expressed by the following formula (\ref{formula10}).

\begin{equation}
	{C_i} = \sigma \left( {LN\left( {conv1{D_i}\left( {ROI{s_i}} \right)} \right)} \right)
	\label{formula10}
\end{equation}
where $ROI{s_i}$ represents the feature matrix of the ROIs belonging to the $i$-th community, $LN$ denotes layer normalization, and $\sigma$ is the activation function. It is important to note that since $conv1D$ can only be implemented in Euclidean space, the features of the ROIs used in this process are first transformed from hyperbolic space to Euclidean space using the transformation defined in formula (\ref{formula4}).

\subsubsection{Community-level HGCN and Whole-Brain Representation Extraction}
To closely simulate the message-passing dynamics between functional brain communities, we proposed a community-level functional network implemented through Hyperbolic Graph Convolutional Networks, which effectively enables information exchange across communities and facilitates the exploration of their complex interrelationships. For the definition of the intricate connection structure within the community network, we refer to the approach proposed in~\cite{Ding2024LGGNet}, where the interaction between each community and the others is modeled using the dot product of their respective features, thereby capturing the functional relationships between communities. This connection model is both dynamic and instance-specific. We assumed that global connections were undirected, as the relationship between two local graphs was inherently reciprocal. Therefore, the adjacency matrix ${A_{C-base}} \in {R^{k \times k}}$ of the community graph is defined as follows:

\begin{equation}
	{A_{C-base}} = \left( {\begin{array}{*{20}{c}}
			{{C_1} \cdot {C_1}}& \ldots &{{C_1} \cdot {C_k}}\\
			\vdots & \ddots & \vdots \\
			{{C_1} \cdot {C_k}}& \cdots &{{C_k} \cdot {C_k}}
	\end{array}} \right)
	\label{formula11}
\end{equation}
Here, $\cdot$ stands for the dot product.

To better capture and learn the complex relationships between functional communities, this study employs a symmetric, trainable attention matrix to emphasize the important connections in the similarity adjacency matrix. It is important to note that the attention matrix is also symmetric, as the adjacency matrix is undirected. The attention matrix $M_C$ is defined as follows:

\begin{equation}
	{M_C} = \left( {\begin{array}{*{20}{c}}
			{{w_{1,1}}}& \ldots &{{w_{1,k}}}\\
			\vdots & \ddots & \vdots \\
			{{w_{k,1}}}& \cdots &{{w_{k,k}}}
	\end{array}} \right)
	\label{formula12}
\end{equation}

Therefore, the complete construction of the functional community network adjacency matrix is given by the following formula (\ref{formula13}):

\begin{equation}
	{A_C} = {\sigma _{relu}}\left( {{A_{C-base}} \circ {M_C}} \right) + I
	\label{formula13}
\end{equation}
where $\circ$ represents the Hadamard product, and $I$ denotes the identity matrix. The inclusion of the identity matrix ensures that the adjacency matrix has self-connections. By adding the identity matrix after the activation function, we can emphasize the strength of self-connections.

It is important to note that the obtained community adjacency matrix $A_C$ differs from the ROI adjacency matrix $A_R$. $A_R$ has already been normalized and then sparsified within the range of $\left[-1, 1\right]$. However, $A_C$ has not undergone normalization. Therefore, before inputting the community adjacency matrix $A_C$ into the community HGCN, the normalization operation is required. The normalized community adjacency matrix can be computed using the following formula:
\begin{equation}
	{\widetilde A_C} = {\widetilde D^{ - \frac{1}{2}}}{A_C}{\widetilde D^{ - \frac{1}{2}}}
	\label{formula14}
\end{equation}
where $\widetilde D$ denotes the degree matrix of $A_C$.

At this point, the corresponding functional community network graph ${G_C}\left( {{V_C},{{\widetilde A}_C}} \right)$ can be constructed and input into a single layer of HGCN for message passing and feature transformation between the community networks.

\begin{equation}
	{H_C} = \exp _0^{\kappa}\left( {\sigma \left( {{{\widetilde A}_C}\left( {\log _0^{\kappa}\left( {HLinear\left( {{V_C},{W_C}} \right)} \right)} \right)} \right)} \right)
	\label{formula15}
\end{equation}

To further compress and represent the features, we designed a whole-brain-scale convolutional layer to achieve a comprehensive representation of the entire brain. This Conv2D uses kernels with the same size as the community networks to capture global brain relationships. To learn the whole-brain relationships from multiple perspectives, we employed multiple kernels (32 kernels in total). For effective Conv2D operation, the features are first mapped to Euclidean space using the logarithmic map as shown in formula (\ref{formula4}) before applying the convolution.

Following the whole-brain-scale convolution, we obtain the feature vector representation for each subject. A single $HLinear$ layer is subsequently employed to perform the final classification. To optimize the model parameters, the cross-entropy loss function (\ref{formula16}) is utilized to guide the training process and minimize the classification error.

\begin{equation}
	{L_{CE}} = \frac{1}{N}\sum\limits_i^N { - \left[ {{y_i} \times \log \left( {{p_i}} \right) + \left( {1 - {y_i}} \right) \times \log \left( {1 - {p_i}} \right)} \right]} 
	\label{formula16}
\end{equation}

\subsubsection{Hyperbolic Space Domain Alignment and Prototype Learning}
For diagnostic classification across multiple sites, especially when the target site data lacks labeled annotations, directly applying a model trained on the source domain to predict target domain data often leads to suboptimal performance due to domain shift and data heterogeneity between the sites.

Maximum Mean Discrepancy is a non-parametric method used to measure the discrepancy between two distributions $P$ and $Q$. The core idea of MMD is to compare the distances between the means of these distributions after mapping them into a Reproducing Kernel Hilbert Space defined by a Gaussian kernel~\cite{Yan2020Weighted}. In the context of hyperbolic space (Poincar$\acute{e}$ disk model), both the data distribution and the distance metric differ from Euclidean space. Therefore, MMD needs to be redefined to align with the characteristics of hyperbolic geometry. We propose the Hyperbolic MMD (HMMD), which is based on the hyperbolic distance metric, and the Gaussian kernel in hyperbolic space is defined as follows:

\begin{equation}
	{K_G}\left( {u,v} \right) = \exp \left( { - \frac{{d\left( {u,v} \right)}}{\gamma }} \right)
	\label{formula17}
\end{equation}

To further enhance the expressiveness of the kernel, we employ a multi-bandwidth Gaussian kernel, defined as:

\begin{equation}
	{K_G}\left( {u,v} \right) = \sum\limits_{i = 1}^L {\exp \left( { - \frac{{d\left( {u,v} \right)}}{{{\gamma _i}}}} \right)} 
	\label{formula18}
\end{equation}
where $L$ represents the number of Gaussian kernels, and $\gamma_i$ denotes the bandwidth for each kernel.

Given a source domain sample $\left\{ {{x_i}} \right\}_{i = 1}^n \sim S$ and a target domain sample $\left\{ {{y_j}} \right\}_{j = 1}^m \sim {\rm T}$, the HMMD can be represented as:
\begin{align}
	&HMMD\left( {S,T} \right) = {E_{x,x' \sim S}}\left[ {{K_G}\left( {x,x'} \right)} \right] \notag \\ &\quad + {E_{y,y' \sim {\rm T}}}\left[ {{K_G}\left( {y,y'} \right)} \right] - 2{E_{x \sim S,y \sim T}}\left[ {{K_G}\left( {x,y} \right)} \right]
	\label{formula19}
\end{align}
$E$ denotes the mathematical expectation. Here $x$ and $x'$ are two independent samples drawn from distribution $S$, while $y$ and $y'$ are independently sampled from distribution $T$.

Unlike standard MMD, which computes distribution discrepancy in Euclidean space, HMMD performs domain alignment in hyperbolic space, leveraging hyperbolic distance to better capture hierarchical structures. This adaptation enables low distortion alignment of complex brain network features~\cite{Ganea2018Hyperbolic}, which are inherently non-Euclidean in nature. To achieve better alignment, as shown in Figure \ref{model_fig}, we perform HMMD alignment on the features before and after the community graph network, as well as on the final representation vectors for feature alignment.

The proposed HMMD alignment achieves marginal distribution alignment. Inspired by~\cite{TANG2022Multitarget}, we introduced prototype alignment to achieve conditional distribution alignment. Specifically, given all points ${X_c} = \left\{ {{x_1},{x_2},{x_3} \ldots ,{x_{{N_c}}}} \right\}$ of a particular class $c$, where $N_c$ denotes the total number of samples in class $c$, the first point is selected as the base point $x_{base}$. The relative displacement of each point with respect to the base point is then computed. The average of all such relative displacements is subsequently calculated. Finally, the averaged relative displacement is mapped back to $x_{base}$, resulting in the class prototype, as shown in formula (\ref{formula20}).

\begin{equation}
	P{t_c} = {x_{base}} \oplus \frac{1}{{{N_c}}}\sum\limits_{i = 1}^{{N_c}} {\left( { - {x_{base}} \oplus {x_i}} \right)} 
	\label{formula20}
\end{equation}

Since the target domain lacks labeled data, we utilized the predicted labels (pseudo-labels) from the alignment step as surrogate ground truth labels to compute the corresponding prototypes. Finally, the hyperbolic distance between the prototypes of the same class in both domains is calculated using formula (\ref{formula21}), leading to the Prototype Alignment (PA) Loss.

\begin{equation}
	{L_{PA}} = \sum\limits_{c = 1}^{{N_{class}}} {d\left( {Pt_c^{source},Pt_c^{target}} \right)} 
	\label{formula21}
\end{equation}

Therefore, our final objective function can be expressed as $L_{total}$:

\begin{align}
	{L_{total}} = &{L_{CE}} + {\lambda _m}\left( {HMM{D_1} + HMM{D_2} + HMM{D_3}} \right) \notag \\ &\quad + {\lambda _{PA}}{L_{PA}}
	\label{formula22}
\end{align}
where $L_{CE}$ represents the cross-entropy loss, $HMMD_{i}$ refers to the marginal distribution alignment loss at different layers, and $L_{PA}$ denotes the conditional distribution alignment loss (Prototype Alignment Loss). $\lambda _m$ and $\lambda _{PA}$ are the weight hyperparameters associated with $HMMD_{i}$ and $L_{PA}$, respectively.

During the testing phase, we obtained the representation vectors from the test data, and then computed the distance between the test vectors and the source domain class prototypes to predict the class of the test data. Here, we did not incorporate the prototypes generated from the target domain data, as the pseudo-labels of the target domain were relatively unstable and may introduce instability during the testing phase.

\section{Experimental Design and Results Analysis}
\subsection{Experimental Setup}
\begin{figure}[htbp] 
	\vspace{-0.6em}
	\centering
	\includegraphics[width=\linewidth]{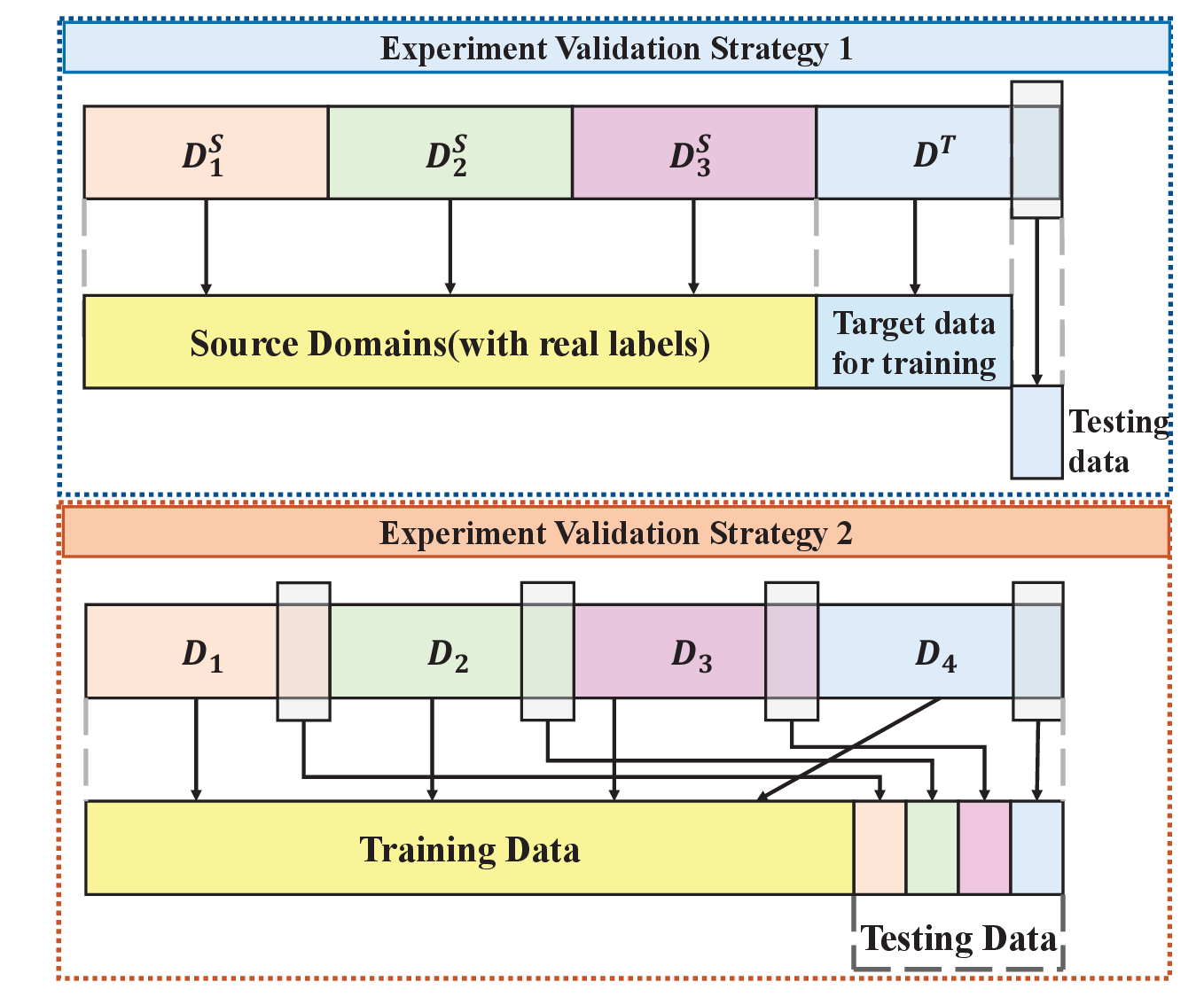}
	\caption{The schematic diagram of the two experimental validation strategies.}
	\label {valid_split}
	\vspace{-0.6em}
\end{figure}

Two experimental validation strategies were employed to evluate the proposed model  H$^2$MSDA during the experiments. In the first validation strategy, each site was used sequentially as the target domain. The data from the current target domain was split into 5 folds, with one fold used as the test set and the remaining four folds as the training data (without real labels) for domain alignment. The data from the other three sites (with real labels) served as the source domain for training. For the second validation strategy, we aimed to assess the reliability of the baseline model by evaluating its performance without domain alignment. Specifically, all site data were combined and randomly shuffled, and classification was performed directly on the mixed data without any domain adaptation. A 5-fold cross-validation was then conducted on the merged dataset. These two experimental validation strategies are illustrated in Figure \ref{valid_split}.

For evaluating the classification performance of the model, we utilized four metrics, i.e., accuracy, specificity, sensitivity and F1 score. The corresponding formulas for these metrics are listed as follows:

\begin{equation}
	Acc = \frac{{TP + TN}}{{TP + FN + FP + TN}}
	\label{ACC}
\end{equation}

\begin{equation}
	Sen = \frac{{TP}}{{TP + FN}}
	\label{SEN}
\end{equation}

\begin{equation}
	Spe = \frac{{TN}}{{TN + FP}}
	\label{SPEC}
\end{equation}

\begin{equation}
	F1 = \frac{{2TP}}{{2TP + FP + FN}}
	\label{F1}
\end{equation}
where TP, TN, FP, and FN denote True Positives, True Negatives, False Positives, and False Negatives, respectively.

\subsection{Compared Methods}
In the experiment, we compared our H$^2$MSDA method with the following approaches, which include seven domain adaptation methods and six non-domain adaptation methods.

The compared domain adaptation methods including:

1) \textbf{A$^2$GCN}~\cite{chu2022resting}\textbf{:}
A$^2$GCN employs an attention-augmented Graph Convolutional Network for domain adaptation in resting-state fMRI. It transfers knowledge from a labeled source to an unlabeled target domain, with attention mechanisms assigning adaptive importance to brain regions.

2) \textbf{Fed\_Align}~\cite{LI2020Multisite}\textbf{:}
Fed\_Align introduces a federated learning-based approach to align feature distributions across multi-site fMRI datasets. It ensures domain alignment during model training and aggregation to improve cross-site consistency.

3) \textbf{TMJDA}~\cite{zhang2020transport}\textbf{:}
TMJDA integrates optimal transport with joint distribution alignment for ASD classification across multiple sites, enabling more precise modeling of functional connectivity under domain shifts.

4) \textbf{maLRR}~\cite{Wang2020Identifying}\textbf{:}
maLRR leverages low-rank regression to learn a shared subspace across sites, mitigating domain discrepancies and enhancing ASD classification on multi-site fMRI data.

5) \textbf{Fed\_MOE}~\cite{LI2020Multisite}\textbf{:}
Fed\_MOE incorporates a mixture-of-experts model into federated learning, where each expert specializes in a site. A global selector dynamically assigns site data to the most suitable expert to enhance domain adaptation.

6) \textbf{TP-MIDA}~\cite{Kunda2023Improving}\textbf{:}
TP-MIDA employs Tangent Person Embedding with maximum independence domain adaptation to reduce site-specific bias while preserving relevant features for autism classification.

7) \textbf{LRCDR}~\cite{Liu2023DomainA}\textbf{:}
LRCDR combines low-rank representation learning with class-discriminative regularization to align distributions across sites and improve ASD classification performance.

The compared non-domain adaptation methods including:

1) \textbf{f-GCN}~\cite{jiang2020hi}\textbf{:}
f-GCN is a threshold-based graph convolutional model that prunes redundant connections (optimal threshold: 0.3). It stacks multiple GCN layers followed by global average pooling and readout to generate compact graph-level representations for classification.

2) \textbf{BrainNetCNN}~\cite{kawahara2017brainnetcnn}\textbf{:}
BrainNetCNN is tailored for brain network analysis, employing Edge-to-Edge, Edge-to-Node, and Node-to-Graph convolutions to hierarchically extract features and model complex neural interactions for brain disorder prediction.

3) \textbf{Hi-GCN}~\cite{jiang2020hi}\textbf{:}
Hi-GCN constructs a population graph where each node represents a subject, characterized by features extracted via f-GCN. Edges are defined based on pairwise correlation coefficients. A four-layer residual GCNs is then applied to enable inter-subject feature propagation, producing discriminative representations for the final classification task.

4) \textbf{MVS-GCN}~\cite{wen2022mvs}\textbf{:}
MVS-GCN is a multi-view GCN model that fuses structural and functional connectivity under prior brain structure guidance. It enhances ASD diagnosis by capturing complementary and biologically relevant brain features.

5) \textbf{MCG-Net}~\cite{luo2023aided}\textbf{:}
MCG-Net adopts a dual-branch architecture that combines convolutional and graph convolutional layers to extract high and low order features. The fused representations are used for classification of brain functional networks.

6) \textbf{ASD-HNet}~\cite{luo2024hierarchical}\textbf{:}
ASD-HNet performs hierarchical feature extraction guided by functional gradients and community structure. It builds an interpretable diagnostic model by combining structural hierarchy with neural network learning for ASD classification.

\begin{table*}[]
	\renewcommand{\arraystretch}{1.3}
	\caption{Performance of our proposed model compared with state-of-the-art methods on NYU and UM target sites. The best results are presented in bold. '$^\dag$' indicates that the results for this method are derived from the original study, while '-' denotes that the experimental results are not reported for this dataset. The superscript symbols ($^*$, $^{**}$, $^{***}$) indicate statistically significant differences compared with H$^2$MSDA, based on paired $t$-test ($^*p<0.05$, $^{**}p<0.01$, $^{***}p<0.001$).}
	\label{NYU_UM_result}
	\centering
	\resizebox{\textwidth}{!}{
	\begin{tabular}{clcccccccccccl}
		\hline
		\multicolumn{1}{l}{\multirow{2}{*}{Group}} & \multirow{2}{*}{Model} & \multicolumn{4}{c}{NYU}                                                                 & \multicolumn{1}{l}{} & \multicolumn{4}{c}{UM}                                                                 &                      & \multirow{2}{*}{Altas} &  \\ \cline{3-6} \cline{8-12}
		\multicolumn{1}{l}{}                       &                        & ACC                 & SEN                  & SPE                  & F1                  &                      & ACC                 & SEN                  & SPE                 & F1                  &                      &                        &  \\ \hline
		\multirow{6}{*}{Non-domain Adaptation}     & f-GCN                  & 71.95±7.55$^*$      & 58.07±24.02$^{*}$    & 79.95±11.75          & 60.89±19.76$^{*}$   &                      & 75.30±6.44${^*}$    & \textbf{82.29±16.97} & 70.25±17.47$^{***}$ & 73.14±7.39          &                      & AAL                    &  \\
		& BrainNetCNN            & 72.52±5.26          & 63.71±33.71          & 78.53±20.29          & 62.96±25.90         &                      & 74.27±8.03          & 64.81±38.13          & 77.53±25.11         & 59.73±34.63         &                      & AAL                    &  \\
		& Hi-GCN                 & 73.65±4.84$^{***}$  & 73.65±4.84$^{**}$    & 68.87±11.01          & 68.29±9.78$^{**}$   &                      & 77.98±10.25         & 77.98±32.41          & 65.19±32.41         & 67.00±25.66         & \multicolumn{1}{l}{} & AAL                    &  \\
		& MVS-GCN                & 74.87±4.75$^*$      & 74.87±4.75           & 53.01±15.77$^{***}$  & 71.78±5.84          &                      & 70.87±12.81$^{**}$  & 70.87±12.81          & 52.14±26.99$^{**}$  & 64.44±12.82         &                      & AAL                    &  \\
		& MCG-Net                & 71.98±7.02$^*$      & 54.22±22.66$^{*}$    & 83.05±11.47          & 59.59±19.08$^{*}$   &                      & 78.77±13.33         & 70.86±36.09          & 81.73±13.77         & 68.94±27.71         &                      & AAL                    &  \\
		& ASD-HNet               & 76.62±8.51          & 71.97±5.89           & 81.11±15.67          & 71.88±7.34          &                      & 77.83±10.80         & 83.95±11.33          & 74.01±10.45$^{*}$   & 75.30±12.68         &                      & AAL                    &  \\ \hline
		\multirow{8}{*}{Domain Adaptation}         & A$^2$GCN               & 66.67±5.27$^{***}$  & 41.96±37.64$^{**}$   & 81.64±19.60          & 41.93±33.44$^{**}$  &                      & 68.18±4.41$^{***}$  & 65.52±33.84          & 68.43±26.48$^{*}$   & 58.67±18.99$^{*}$   &                      & AAL                    &  \\
		& Fed\_Align             & 71.97±8.31          & 51.51±13.77$^{***}$  & \textbf{84.64±11.83} & 59.25±10.76$^{**}$  &                      & 76.33±3.76$^{**}$   & 45.87±7.36$^{***}$   & \textbf{91.67±7.76} & 53.11±3.69$^{***}$  &                      & AAL                    &  \\
		& TMJDA$^\dag$           & 71.09               & -                    & -                    & -                   &                      & 72.57               & 75.00                & 70.77               & -                   &                      & AAL                    &  \\
		& maLRR$^\dag$           & 71.88±4.42          & 66.67±3.57           & 78.57±10.20          & -                   &                      & 72.73±6.43          & 76.92±1.76           & 66.67±7.14          & -                   &                      & AAL                    &  \\
		& Fed\_MOE               & 67.83±8.56$^{**}$   & 73.45±20.83          & 64.64±11.88$^{*}$    & 65.56±18.06$^{*}$   &                      & 70.07±12.31${^*}$   & 47.10±11.54$^{**}$   & 84.92±8.52          & 55.51±11.49$^{***}$ &                      & AAL                    &  \\
		& TP-MIDA$^\dag$         & 75.60               & 75.10                & -                    & 75.00               &                      & 72.40               & 73.60                & -                   & 71.40               &                      & CC200                  &  \\
		& LRCDR$^\dag$           & 73.40               & 78.50                & 69.50                & 67.70               &                      & 74.50               & 70.60                & 77.90               & \textbf{75.80}      &                      & AAL                    &  \\
		& H$^2$MSDA(ours)        & \textbf{80.66±5.39} & \textbf{82.72±12.76} & 78.20±5.71           & \textbf{78.01±5.71} &                      & \textbf{81.42±1.81} & 73.42±22.57          & 83.40±13.48         & 74.72±13.48         &                      & AAL                    &  \\ \hline
		\end{tabular}}
\end{table*}

\begin{table*}[]
	\renewcommand{\arraystretch}{1.3}
	\caption{Performance of our proposed model compared with state-of-the-art methods on USM and UCLA target sites. The best results are presented in bold. '$^\dag$' indicates that the results for this method are derived from the original study, while '-' denotes that the experimental results are not reported for this dataset. The superscript symbols ($^*$, $^{**}$, $^{***}$) indicate statistically significant differences compared with H$^2$MSDA, based on paired $t$-test ($^*p<0.05$, $^{**}p<0.01$, $^{***}p<0.001$).}
	\label{USM_UCLA_result}
	\centering
	\resizebox{\textwidth}{!}{
	\begin{tabular}{llcccccccccccl}
		\hline
		\multirow{2}{*}{Group}                 & \multirow{2}{*}{Model} & \multicolumn{4}{c}{USM}                                                                & \multicolumn{1}{l}{} & \multicolumn{4}{c}{UCLA}                                                     &                      & \multirow{2}{*}{Altas} &  \\ \cline{3-6} \cline{8-12}
		&                        & ACC                 & SEN                  & SPE                 & F1                  &                      & ACC                 & SEN            & SPE                  & F1             &                      &                        &  \\ \hline
		\multirow{6}{*}{Non-domain Adaptation} & f-GCN                  & 80.12±9.70$^*$      & 70.62±18.31$^{**}$   & 86.66±29.81         & 79.61±13.87$^{*}$   &                      & 80.00±6.66$^{**}$   & 59.21±34.97    & 89.14±17.42          & 64.71±36.57    &                      & AAL                    &  \\
		& BrainNetCNN            & 78.59±7.68$^{**}$   & 83.42±12.83          & 69.62±9.71          & 81.71±5.79$^{*}$    &                      & 70.67±3.64$^{***}$  & 47.36±16.72    & 86.43±12.39          & 52.39±10.54    &                      & AAL                    &  \\
		& Hi-GCN                 & 86.79±11.01         & 86.79±6.48           & \textbf{89.56±6.48} & 88.49±9.53          &                      & 80.00±8.16$^{**}$   & 80.00±16.86    & 71.07±16.86$^{***}$  & 76.38±7.84     & \multicolumn{1}{l}{} & AAL                    &  \\
		& MVS-GCN                & 78.59±3.60$^{***}$  & 78.59±3.60$^{***}$   & 77.20±11.28         & 76.33±5.20$^{***}$  &                      & 69.33±3.65$^{***}$  & 69.33±3.65     & 44.86±25.34$^{***}$  & 62.54±11.62    &                      & AAL                    &  \\
		& MCG-Net                & 78.59±9.68$^*$      & 80.75±19.76          & 79.76±16.24         & 82.01±13.86         &                      & 70.67±5.96$^{***}$  & 47.57±33.21    & 84.50±17.53          & 52.41±31.90    &                      & AAL                    &  \\
		& ASD-HNet               & 83.72±4.28$^{***}$  & 88.61±10.17          & 76.48±16.08         & 87.29±3.59          &                      & 69.33±11.54$^{***}$ & 64.36±19.93    & 75.16±16.16          & 65.26±10.55    &                      & AAL                    &  \\ \hline
		\multirow{8}{*}{Domain Adaptation}     & A$^2$GCN               & 73.72±7.13$^{***}$  & 74.26±18.65$^{**}$   & 65.29±38.13         & 75.98±12.10$^{**}$  &                      & 69.33±10.11$^{***}$ & 57.50±32.98    & 73.86±42.05          & 59.11±21.86    &                      & AAL                    &  \\
		& Fed\_Align             & 67.44±10.95$^{***}$ & 61.33±17.05$^{***}$  & 79.17±14.91         & 69.47±16.12$^{***}$ &                      & 72.00±10.11$^{***}$ & 51.79±12.77    & 90.83±16.41          & 61.14±11.99    &                      & AAL                    &  \\
		& TMJDA$^\dag$           & 80.18               & -                    & -                   & -                   &                      & 73.10               & -              & -                    & -              &                      & AAL                    &  \\
		& maLRR$^\dag$           & 74.62±5.32          & 86.11±2.73           & 68.10±4.30          & -                   &                      & 75.00±5.05          & 64.29±1.01     & 85.71±2.00           & -              &                      & AAL                    &  \\
		& Fed\_MOE               & 73.72±9.75$^{***}$  & \textbf{97.50±11.67} & 32.50±12.44$^{***}$ & 80.30±10.59         &                      & 65.33±16.44$^{***}$ & 82.44±10.59    & 45.01±30.61$^{***}$  & 67.27±29.61    &                      & AAL                    &  \\
		& TP-MIDA$^\dag$         & 79.30               & 80.70                & -                   & 77.80               &                      & 82.60               & \textbf{82.80} & -                    & \textbf{81.70} &                      & CC200                  &  \\
		& LRCDR$^\dag$           & 74.30               & 75.90                & 72.10               & 75.20               &                      & 71.70               & 70.40          & 73.30                & 74.60          &                      & AAL                    &  \\
		& H$^2$MSDA(ours)        & \textbf{88.46±4.73} & 94.17±7.37           & 76.00±9.93          & \textbf{91.11±5.66} &                      & \textbf{89.33±3.65} & 71.03±40.53    & \textbf{91.11±14.48} & 72.76±40.72    &                      & AAL                    &  \\ \hline
		\end{tabular}}
\end{table*}

\subsection{Result on Multi-Site data}
In this section, we conducted experimental comparisons between the proposed H$^2$MSDA and other methods. The experimental results for the two different experiment validation strategies are presented and analyzed as follows.

\subsubsection{Results on Experiment Validation Strategy 1}
Each site was sequentially selected as the target domain, while the remaining three sites served as the source domain. It is important to note that since the comparison includes non-domain adaptation methods, we employed 5-fold cross-validation on the data from each site for non-domain adaptation methods to ensure a fair comparison. In simpler terms, the unlabeled target domain data used for training in domain adaptation methods was instead used with labels for training in non-domain adaptation methods. The experimental results across the four sites are presented in Tables \ref{NYU_UM_result} and \ref{USM_UCLA_result}.

From the experimental results, it is evident that the proposed method achieves superior performance in terms of accuracy across all four sites. Compared to the baseline methods, our approach demonstrates improvements of +4.02\%, +2.65\%, +1.67\%, and +6.73\% for NYU, UM, USM, and UCLA, respectively. While our method may not outperform others on every metric, it achieves a balanced performance in terms of specificity and sensitivity, avoiding extreme distributions (e.g., very low sensitivity paired with very high specificity). Furthermore, we conducted paired $t$-test to assess the statistical significance on the performance among different models. As indicated in the results Tables \ref{NYU_UM_result} and \ref{USM_UCLA_result}, most compared methods show statistically significant differences when compared to our model, while several methods exhibit nonsignificant differences. This highlights the robustness and reliability of our approach, ensuring a more consistent performance across different metrics.

\begin{figure}[htbp] 
	\vspace{-0.6em}
	\centering
	\includegraphics[width=\linewidth]{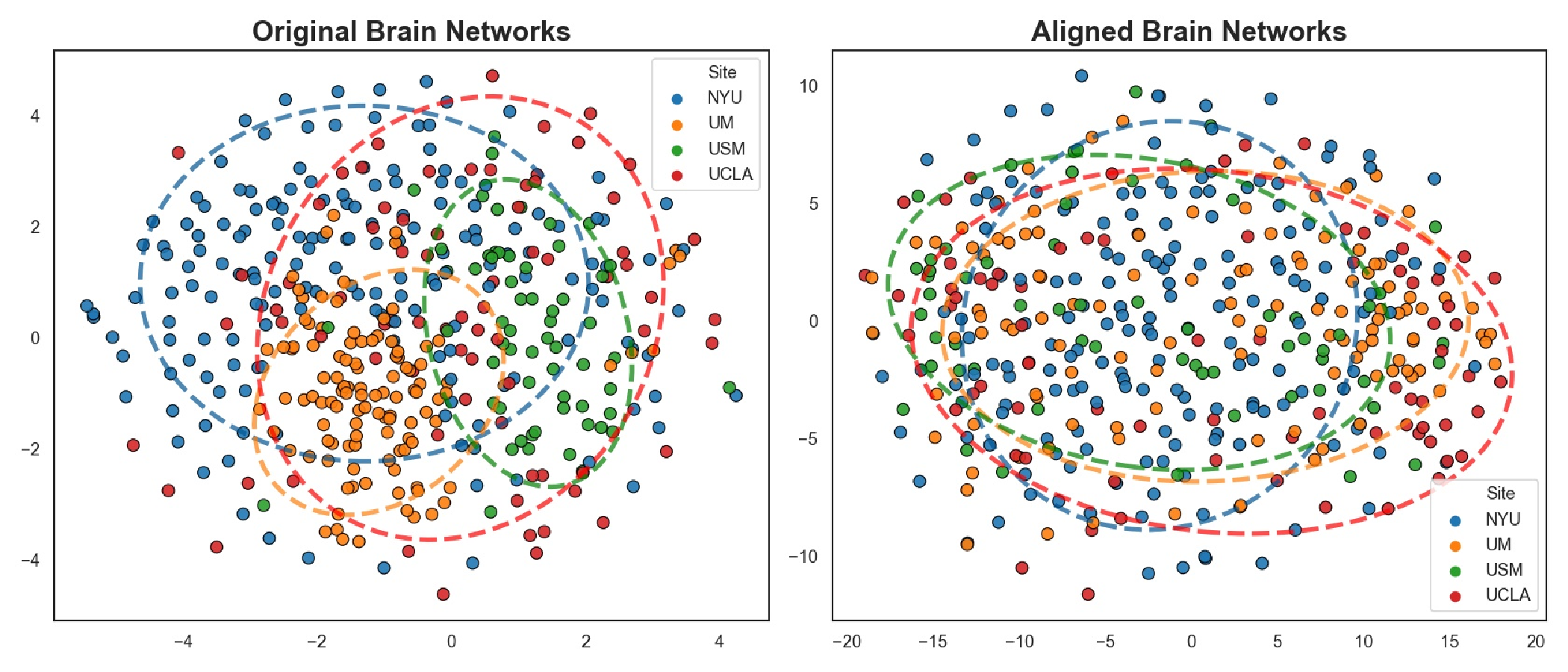}
	\caption{Visualization of multi-site data distribution for the original and aligned brain networks in a 2D space via $t$-SNE. Each point represents a brain network and different colors indicate different sites. To provide a more intuitive observation of the distribution variation for different sites, we draw ellipses for the 4 sites.}
	\label {tsne}
	\vspace{-0.6em}
\end{figure}

To visualize the data distribution of brain networks before and after domain adaptation, we applied $t$-SNE~\cite{SONG2024BrainDAS} to map the original brain networks and the extracted representation networks into a two-dimensional space. As shown in Figure \ref{tsne}, each point in the $t$-SNE scatter plot represents a sample of a brain network. The visualization clearly demonstrates that, after applying the alignment operation, the data distribution across different sites becomes more consistent, with no apparent domain shift. This result validates the effectiveness of the proposed method in addressing cross-domain heterogeneity.

\begin{table}[]
	\renewcommand{\arraystretch}{1.2}
	\caption{Results on Experiment Validation Strategy 2. The best results are presented in bold.}
	\label{EES2_results}
	\begin{tabular}{lllcccccccl}
		\hline
		& Methods         &  & ACC            &  & SEN            &  & SPE            &  & F1             &  \\ \hline
		& f-GCN           &  & 65.71          &  & 54.76          &  & 73.99          &  & 59.42          &  \\
		& BrainNetCNN     &  & 67.62          &  & 57.45          &  & 76.46          &  & 61.96          &  \\
		& Hi-GCN          &  & 66.65          &  & 66.65          &  & 65.24          &  & 63.24          &  \\
		& MVS-GCN         &  & 64.56          &  & 64.56          &  & 52.49          &  & 63.14          &  \\
		& MCG-GCN         &  & 68.34          &  & 66.29          &  & 69.20          &  & 65.88          &  \\
		& ASD-HNet        &  & 71.17          &  & \textbf{67.19} &  & 74.71          &  & 68.14          &  \\
		& H$^2$MSDA(ours) &  & \textbf{72.85} &  & 65.35          &  & \textbf{78.98} &  & \textbf{68.31} &  \\ \hline
	\end{tabular}
\end{table}

\begin{table*}
	\renewcommand{\arraystretch}{1.3}
	\caption{Ablation experiments on NYU site. '-' indicates that the loss is not used. '$\checkmark$' denotes that the use of the loss}
	\centering
	\label{ablation}
	\begin{tabular}{lcccclclclclcl}
		\hline
		& HMMD Loss1                         & HMMD Loss2                         & HMMD Loss3                         & Prototype Align                                         &  & ACC                                &  & SEN                                &  & SPE                       &  & F1                                 &  \\ \hline
		& \textbf{-}                         & \textbf{-}                         & \textbf{-}                         & \multicolumn{1}{c|}{\textbf{-}}                         &  & 74.22                              &  & 61.58                              &  & \textbf{84.09}            &  & 65.39                              &  \\
		& \textbf{\checkmark} & \textbf{\checkmark} & \textbf{\checkmark} & \multicolumn{1}{c|}{\textbf{-}}                         &  & 78.94                              &  & 72.12                              &  & 83.58                     &  & 73.96                              &  \\
		& \textbf{-}                         & \textbf{-}                         & \textbf{-}                         & \multicolumn{1}{c|}{\textbf{\checkmark}} &  & 77.76                              &  & 76.68                              &  & 78.65                     &  & 74.43                              &  \\
		& \textbf{-}                         & \textbf{\checkmark} & \textbf{\checkmark} & \multicolumn{1}{c|}{\textbf{\checkmark}} &  & 80.07                              &  & 78.38                              &  & 80.31                     &  & 76.34                              &  \\
		& \textbf{\checkmark} & \textbf{-}                         & \textbf{\checkmark} & \multicolumn{1}{c|}{\textbf{\checkmark}} &  & \multicolumn{1}{l}{80.07}          &  & \multicolumn{1}{l}{81.05}          &  & \multicolumn{1}{l}{78.20} &  & \multicolumn{1}{l}{76.93}          &  \\
		& \textbf{\checkmark} & \textbf{\checkmark} & \textbf{-}                         & \multicolumn{1}{c|}{\textbf{\checkmark}} &  & \multicolumn{1}{l}{78.53}          &  & \multicolumn{1}{l}{70.90}          &  & \multicolumn{1}{l}{82.12} &  & \multicolumn{1}{l}{72.25}          &  \\
		& \textbf{\checkmark} & \textbf{\checkmark} & \textbf{\checkmark} & \multicolumn{1}{c|}{\textbf{\checkmark}} &  & \multicolumn{1}{l}{\textbf{80.66}} &  & \multicolumn{1}{l}{\textbf{82.72}} &  & \multicolumn{1}{l}{78.20} &  & \multicolumn{1}{l}{\textbf{78.01}} &  \\ \hline
	\end{tabular}
\end{table*}

\subsubsection{Results on Experiment Validation Strategy 2 }
This validation strategy aims to assess the reliability of our baseline model by evaluating its performance in the absence of domain alignment. Specifically, we directly classify using mixed site data without any domain adaptation. The data from all four sites is combined and shuffled, followed by 5-fold cross-validation. The resulting experimental outcomes are presented in Table \ref{EES2_results}. From the results, our baseline model achieved the best performance across multiple metrics, including accuracy (ACC, +1.68\%), specificity (SPE, +2.52\%), and F1-score (+0.17\%). These improvements highlight the robustness and effectiveness of our baseline approach. This experimental evidence provides strong support for the reliability and utility of our baseline model in addressing the classification task effectively.

\subsection{Ablation Experiment}
To validate the effectiveness of the proposed domain alignment constraints, we conducted an ablation study on the corresponding loss functions. Specifically, we selected NYU as the target domain and successively removed each domain alignment loss constraint. The experimental results are presented in Table \ref{ablation}. The results indicate that the model's performance significantly degrades when any domain alignment loss constraint is omitted, with accuracy dropping by 6.44\% in the absence of all constraints. Furthermore, the sequential removal of the four loss terms reveals a consistent decline in performance, with the removal of the prototype alignment loss and HMMD Loss3 resulting in particularly pronounced decreases. Notably, while the full model yields the lowest specificity, it achieves the highest accuracy, sensitivity, and F1-score. This suggests that the alignment enhances the model’s ability to detect positive cases, possibly at the cost of increased false positives, reflecting a trade-off that benefits overall classification performance. These findings demonstrate the effectiveness of the proposed domain alignment constraints in enhancing model performance.

\begin{table}[H]
	\renewcommand{\arraystretch}{1.3}
	\centering
	\caption{Demographic Information of the Five Sites Involved in the Rest-meta-MDD Dataset.}
	\begin{tabular}{ccccccc}
		\hline
		Site &  & MDD &  & HC  &  & Scanner             \\ \hline
		S20  &  & 282 &  & 251 &  & Siemens Tim Trio 3T \\
		S21  &  & 86  &  & 70  &  & Siemens Tim Trio 3T \\
		S1   &  & 74  &  & 74  &  & Siemens Tim Trio 3T \\
		S25  &  & 89  &  & 63  &  & Siemens Verio 3T    \\
		S8   &  & 75  &  & 75  &  & GE Signa 3T         \\ \hline
	\end{tabular}
	\label{MDD_data}
\end{table}

\subsection{Experiments on model generalization}
In order to investigate the generalization performance of the proposed method, we conducted extra experiments on REST-meta-MDD data (http://rfmri.org/REST-meta-MDD). This dataset comprises resting-state fMRI data collected from multiple sites and provides a valuable benchmark for evaluating model robustness across different domains. Specifically, we selected the five sites with the largest sample sizes from the REST-meta-MDD dataset (S1, S8, S20, S21, and S25), comprising a total of 606 patients diagnosed with MDD and 533 healthy controls. The detailed demographic information of these sites is provided in Table~\ref{MDD_data}. To ensure consistency with our previous experiments, the data were preprocessed using the DPARSF toolbox, and the AAL atlas was employed for brain parcellation.

Similarly, we conducted experiments by treating each site as the target domain in turn, using five-fold cross-validation for evaluation (Validation Strategy 1). The experimental results are presented in Table~\ref{MDD_results}. Across the five sites (S20, S21, S1, S25, and S8), our method achieved accuracy improvements of 1.14\%, 2.58\%, 9.95\%, 5.26\%, and 2.66\%, respectively. Moreover, paired $t$-test revealed that our proposed method exhibited statistically significant improvements over most baseline methods. These results demonstrate that the proposed H$^2$MSDA framework maintains strong performance even on data from a different clinical condition, highlighting its excellent generalization ability and robustness.

\begin{table*}[]
	\renewcommand{\arraystretch}{1.3}
	\caption{Experimental results of H$^2$MSDA on REST-meta-MDD data, with best results shown in bold. The superscript symbols ($^*$, $^{**}$, $^{***}$) indicate statistically significant differences compared with H$^2$MSDA, based on paired $t$-test ($^*p<0.05$, $^{**}p<0.01$, $^{***}p<0.001$).}
	\centering
	\label{MDD}
	\begin{tabular}{clllcccccccl}
		\hline
		\multicolumn{1}{l}{Target site} &                      & Model           &                      & ACC(\%)             &           & SEN(\%)              &           & SPE(\%)              &           & F1(\%)              &                               \\ \hline
		\multirow{4}{*}{Site 20}        &                      & A$^2$GCN        & \multicolumn{1}{c}{} & 60.04±4.74$^{**}$   &           & 35.67±27.81$^{*}$    &           & \textbf{77.95±20.17} &           & 40.53±22.09$^{*}$   &                               \\
		&                      & Fed\_Align      & \multicolumn{1}{c}{} & 62.48±0.78$^{***}$  &           & 47.18±11.12          &           & 76.34±11.59          &           & 53.52±5.67          &                               \\
		&                      & Fed\_MOE        & \multicolumn{1}{c}{} & 64.34±3.68          &           & 49.11±7.62$^{*}$     &           & 77.17±3.91           &           & 56.03±7.99          &                               \\
		&                      & H$^2$MSDA(ours) & \multicolumn{1}{c}{} & \textbf{65.48±1.07} &           & \textbf{59.57±14.16} &           & 70.05±12.66          &           & \textbf{61.05±8.16} &                               \\ \hline
		\multirow{4}{*}{Site 21}        &                      & A$^2$GCN        & \multicolumn{1}{c}{} & 66.73±7.60$^{*}$    &           & 39.40±14.66$^{**}$   &           & 87.87±5.81           &           & 50.01±13.44$^{**}$  &                               \\
		&                      & Fed\_Align      &                      & 72.38±6.88          & \textbf{} & 48.56±10.37$^{***}$  &           & 90.06±9.21           &           & 60.49±8.91$^{**}$   &                               \\
		&                      & Fed\_MOE        & \textbf{}            & 62.76±6.98$^{**}$   &           & 25.65±11.88$^{***}$  & \textbf{} & \textbf{91.37±8.19}  & \textbf{} & 36.76±14.04$^{***}$ &                               \\
		&                      & H$^2$MSDA(ours) &                      & \textbf{74.96±5.93} &           & \textbf{73.29±12.40} &           & 74.48±6.85           &           & \textbf{70.91±8.53} &                               \\ \hline
		\multirow{4}{*}{Site 1}         &                      & A$^2$GCN        & \multicolumn{1}{c}{} & 69.59±5.34$^{***}$  &           & 73.19±15.08$^{*}$    &           & 64.65±22.82$^{*}$    &           & 69.90±6.89$^{**}$   &                               \\
		& \multicolumn{1}{c}{} & Fed\_Align      &                      & 71.06±7.92$^{***}$  &           & 58.59±16.93$^{**}$   &           & \textbf{81.99±14.58} &           & 65.77±13.79$^{**}$  & \multicolumn{1}{c}{}          \\
		& \multicolumn{1}{c}{} & Fed\_MOE        &                      & 61.72±8.30$^{***}$  &           & 63.55±17.53$^{*}$    &           & 59.95±18.74          &           & 61.78±9.86$^{***}$  & \multicolumn{1}{c}{}          \\
		& \multicolumn{1}{c}{} & H$^2$MSDA(ours) &                      & \textbf{81.01±5.43} &           & \textbf{85.42±11.80} &           & 73.84±18.45          &           & \textbf{81.49±5.97} & \multicolumn{1}{c}{}          \\ \hline
		\multirow{4}{*}{Site 25}        &                      & A$^2$GCN        & \multicolumn{1}{c}{} & 67.85±7.74$^{**}$   &           & 50.67±35.84          &           & 77.17±25.96          &           & 50.03±29.14$^{*}$   &                               \\
		& \multicolumn{1}{c}{} & Fed\_Align      &                      & 73.04±4.13$^{**}$   &           & 53.29±24.93          &           & \textbf{84.42±11.07} &           & 58.64±19.03         & \multicolumn{1}{c}{}          \\
		& \multicolumn{1}{c}{} & Fed\_MOE        &                      & 73.00±7.30          &           & \textbf{81.05±9.08}  &           & 68.19±11.07          & \textbf{} & 70.94±8.66          & \multicolumn{1}{c}{\textbf{}} \\
		& \multicolumn{1}{c}{} & H$^2$MSDA(ours) &                      & \textbf{78.30±2.78} &           & 72.35±16.83          &           & 80.18±16.93          &           & \textbf{72.27±8.33} & \multicolumn{1}{c}{}          \\ \hline
		\multirow{4}{*}{Site 8}         &                      & A$^2$GCN        &                      & 65.33±8.37$^{**}$   &           & 48.92±29.18$^{*}$    &           & 78.84±14.22          &           & 52.57±30.36$^{*}$   &                               \\
		&                      & Fed\_Align      &                      & 72.67±2.79          &           & 54.87±14.26$^{***}$  &           & \textbf{88.92±8.47}  &           & 65.45±10.11$^{**}$  &                               \\
		&                      & Fed\_MOE        &                      & 65.33±5.06$^{***}$  &           & \textbf{82.06±6.73}  &           & 49.80±11.06$^{***}$  &           & 70.03±4.69$^{**}$   &                               \\
		&                      & H$^2$MSDA(ours) &                      & \textbf{75.33±2.98} &           & 74.43±14.97          &           & 75.75±17.93          &           & \textbf{74.77±3.24} &                               \\ \hline
	\end{tabular}
	\label{MDD_results}
\end{table*}

\section{Discussion}
In this section, we provide a detailed discussion on the impact of functional gradient components and the number of clustered communities on the model's performance. Additionally, we explore the interpretability of the model and analyze the pathological mechanisms identified by the model.
\subsection{The influence of the hyperparameters settings on the model performance}
\begin{figure*}
	\centering
	\includegraphics[width=\linewidth]{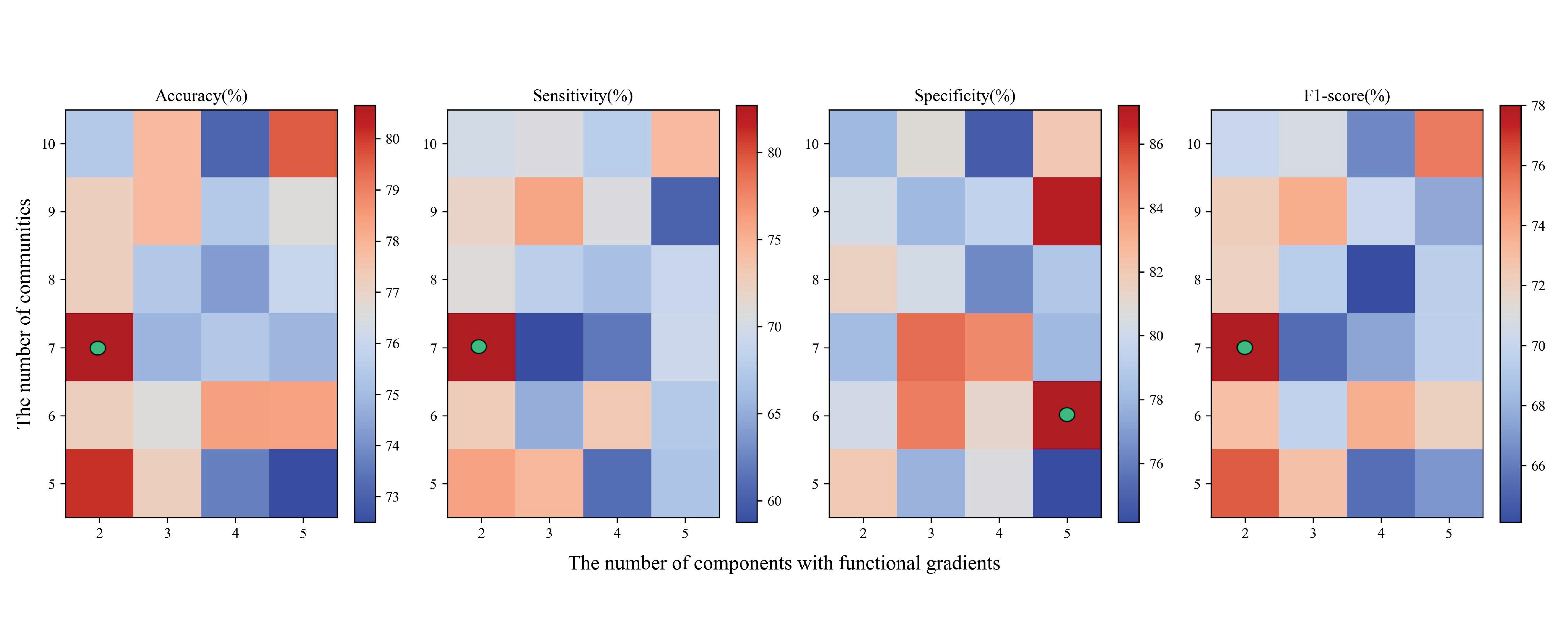}
	\caption{The averaged four metrics obtained by different combination of $k$ and $N_{gra}$. The highest value for each metric is highlighted with a green circle.}
	\label{Search-hyperparameters}
\end{figure*}

The selection of functional gradient components and the number of clustered communities in the model ($N_{gra}$=2, $k$=7) was based on our prior study~\cite{luo2024hierarchical}. However, to ensure that this configuration is both reasonable and effective for the given problem, we conducted a hyperparameter search experiment by exploring different combinations of these parameters. Specifically, we searched for the number of functional gradient components $N_{gra}$ within the range $[2, 3, 4, 5]$ and the number of communities $k$ within the range $[5, 6, 7, 8, 9, 10]$. The results of this search are presented in Figure \ref{Search-hyperparameters}. 

From the results shown in Figure \ref{Search-hyperparameters}, we can conclude that selecting the first two gradient components and seven communities for constructing the community network yields the best performance. The reason for the superior performance of the first two gradient components is consistent with the findings of~\cite{gong2023connectivity}, which highlighted that the first two gradient components account for the majority of variance in the functional connectivity matrix and possess relatively clear physiological significance.

\begin{figure*}
	\centering
	\includegraphics[width=\linewidth]{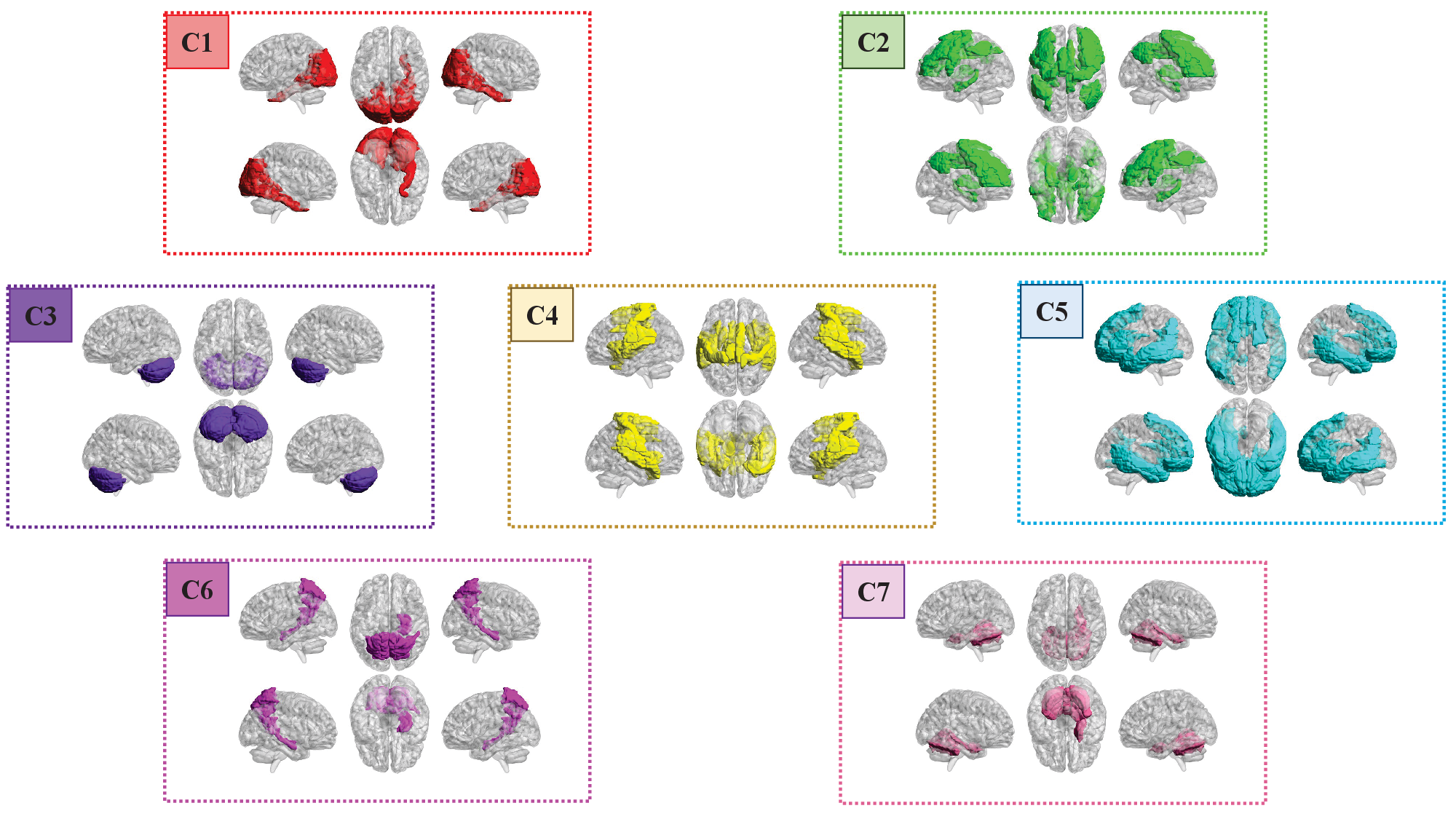}
	\caption{Visualization of the community network results obtained through clustering.}
	\label{comm_results}
\end{figure*}

\subsection{Interpretability}
The results of the community networks obtained through clustering are visually presented in Figure \ref{comm_results}. Through the clustering results, we observe that Community 1 is primarily concentrated in the occipital lobe, which aligns closely with the visual network~\cite{Lee2013Resting}. This network is primarily involved in the processing of visual information. The majority of the brain regions in Community 2 are directly or indirectly involved in the control network, particularly in decision-making, executive control, emotional regulation, and cognitive tasks. For example, regions such as the middle frontal gyrus, insula, and supplementary motor area are directly engaged in cognitive and behavioral control. Meanwhile, areas like the hippocampus and amygdala are more involved in emotional regulation and memory processes, although they also influence control network activities~\cite{chein2005neuroimaging}. Community 4 is predominantly associated with motor control and sensory information processing functions. Community 5 largely overlaps with the default mode network (DMN), which is fundamentally associated with cognitive processes such as introspection, self-reflection, social cognition, and memory recall~\cite{raichle2001default}. Community 6 aligns with the dorsal attention network (DAN), whose activity is typically linked with an individual’s state of awareness, emotional regulation, and perception of the external world~\cite{corbetta2002control}. Communities 3 and 7 primarily consist of the cerebellar hemispheres and the cerebellar vermis, which together form the cerebellum. These regions are crucial for motor control, coordination, and fine-tuning of movements. 

In conclusion, the identified communities are consistent with known brain networks, providing insights into the specific roles of these networks in various cognitive, sensory, and motor functions. The relationship between brain regions within these communities reflects their specialized roles in higher-order cognitive processes and emotional regulation.

\begin{figure}
	\centering
	\includegraphics[width=\linewidth]{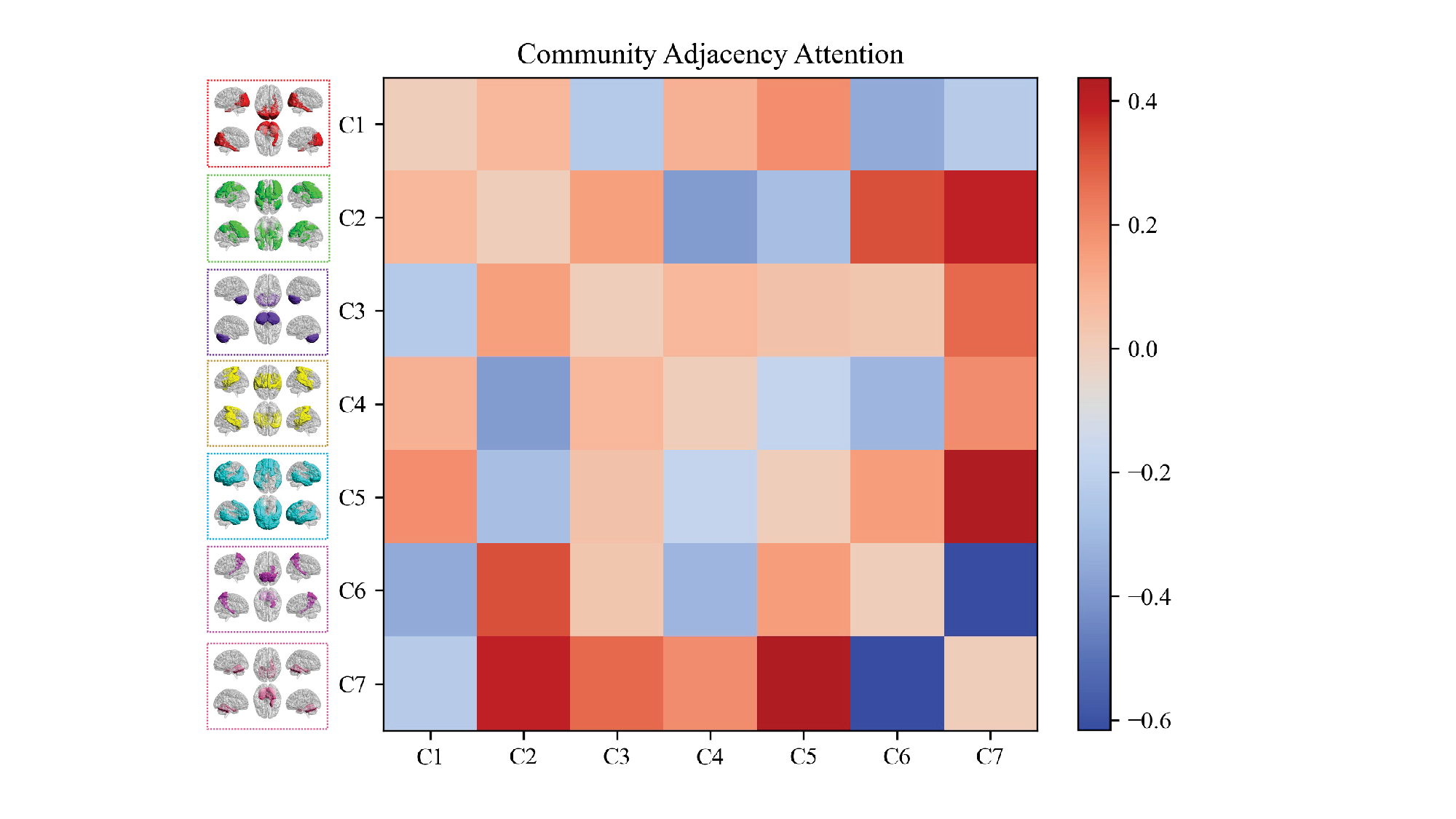}
	\caption{Visualization of the attention mask for the adjacency matrix between communities.}
	\label{comm_atten}
\end{figure}

Additionally, we visualized the attention matrix of the community adjacency matrix for the network described in (\ref{formula12}). To ensure the reliability and stability of the results, we averaged the attention matrices $M_c$ across all folds when using the NYU site as the target domain. Furthermore, we observed that the results for the other three sites as target domains closely resembled those of NYU. Therefore, we present only the visualized attention matrix for the community network adjacency matrix when NYU serves as the target domain, as shown in Figure \ref{comm_atten}.

In this attention weight matrix, values closer to zero indicate weaker connections between two communities. Upon examining the visualization, it is evident that Community 7 exhibits relatively strong attention weights (whether positive or negative) with other communities. This observation aligns well with existing findings~\cite{mosconi2015role,hampson2015autism}, which suggest that the cerebellum plays a crucial role in regulating brain function and has significant associations with ASD.

\subsection{Limitations and Future Work}
 Several limitations should be addressed in future work. First, although we extended our evaluation to include data from multiple sites and both ASD and MDD disorders, the generalizability of the model to other neurological or psychiatric conditions and larger-scale datasets across more diverse populations remains to be further explored. Second, we focused solely on unimodal imaging data, without exploring experiments using multimodal imaging. Extensive research~\cite{xu2021morphological,Wei2023Autistic} has demonstrated that multimodal data can facilitate more efficient diagnoses and improve the identification of abnormal brain connectivity associated with neurological disorders.
 
 In future studies, we aim to incorporate datasets from diverse psychiatric disorders and additional sites to evaluate the generalization and robustness of our method. Furthermore, we plan to leverage multimodal imaging data to enable more comprehensive diagnosis and analysis of brain disorders.
 
\section{Conclusion}
In this paper, we propose H$^2$MSDA model, a novel framework operating in hyperbolic space for multi-site ASD identification. The model performs feature representation learning and prototype-based classification in hyperbolic geometry, enabling more effective modeling of complex brain network structures. To address domain shift, we introduce Hyperbolic Maximum Mean Discrepancy to align marginal distributions across sites. Additionally, a prototype alignment strategy is employed to align class-wise prototypes between source and target domains, thereby achieving conditional distribution alignment. Extensive experiments validate the effectiveness of the proposed method, which outperforms baseline models and demonstrates strong generalization across distinct neurological conditions, including ASD and MDD. These results highlight the model’s potential to support more accurate and robust clinical decision-making in multi-site neuroimaging studies.


%
%

\section*{References}

\bibliographystyle{IEEEtran}
\bibliography{reference.bib}

\begin{thebibliography}{10}
\providecommand{\url}[1]{#1}
\csname url@samestyle\endcsname
\providecommand{\newblock}{\relax}
\providecommand{\bibinfo}[2]{#2}
\providecommand{\BIBentrySTDinterwordspacing}{\spaceskip=0pt\relax}
\providecommand{\BIBentryALTinterwordstretchfactor}{4}
\providecommand{\BIBentryALTinterwordspacing}{\spaceskip=\fontdimen2\font plus
\BIBentryALTinterwordstretchfactor\fontdimen3\font minus
  \fontdimen4\font\relax}
\providecommand{\BIBforeignlanguage}[2]{{%
\expandafter\ifx\csname l@#1\endcsname\relax
\typeout{** WARNING: IEEEtran.bst: No hyphenation pattern has been}%
\typeout{** loaded for the language `#1'. Using the pattern for}%
\typeout{** the default language instead.}%
\else
\language=\csname l@#1\endcsname
\fi
#2}}
\providecommand{\BIBdecl}{\relax}
\BIBdecl

\bibitem{pandolfi2018screening}
V.~Pandolfi, C.~I. Magyar, and C.~A. Dill, ``Screening for autism spectrum
  disorder in children with down syndrome: An evaluation of the pervasive
  developmental disorder in mental retardation scale,'' \emph{Journal of
  Intellectual \& Developmental Disability}, vol.~43, no.~1, pp. 61--72, 2018.

\bibitem{vohra2017comorbidity}
R.~Vohra, S.~Madhavan, and U.~Sambamoorthi, ``Comorbidity prevalence,
  healthcare utilization, and expenditures of medicaid enrolled adults with
  autism spectrum disorders,'' \emph{Autism}, vol.~21, no.~8, pp. 995--1009,
  2017.

\bibitem{lord2018autism}
C.~Lord, M.~Elsabbagh, G.~Baird, and J.~Veenstra-Vanderweele, ``Autism spectrum
  disorder,'' \emph{The lancet}, vol. 392, no. 10146, pp. 508--520, 2018.

\bibitem{hirota2023autism}
T.~Hirota and B.~H. King, ``Autism spectrum disorder: A review,'' \emph{Jama},
  vol. 329, no.~2, pp. 157--168, 2023.

\bibitem{Ye2024MAD}
J.~Ye, A.~Zeng, D.~Pan, Y.~Zhang, J.~Zhao, Q.~Chen, and Y.~Liu, ``Mad-former: A
  traceable interpretability model for alzheimer's disease recognition based on
  multi-patch attention,'' \emph{IEEE Journal of Biomedical and Health
  Informatics}, vol.~28, no.~6, pp. 3637--3648, 2024.

\bibitem{matthews2004functional}
P.~M. Matthews and P.~Jezzard, ``Functional magnetic resonance imaging,''
  \emph{Journal of Neurology, Neurosurgery \& Psychiatry}, vol.~75, no.~1, pp.
  6--12, 2004.

\bibitem{friston1994analysis}
K.~J. Friston, P.~Jezzard, and R.~Turner, ``Analysis of functional mri
  time-series,'' \emph{Human brain mapping}, vol.~1, no.~2, pp. 153--171, 1994.

\bibitem{woodward2015resting}
N.~D. Woodward and C.~J. Cascio, ``Resting-state functional connectivity in
  psychiatric disorders,'' \emph{JAMA psychiatry}, vol.~72, no.~8, pp.
  743--744, 2015.

\bibitem{FANG2025Source}
\BIBentryALTinterwordspacing
Y.~Fang, J.~Wu, Q.~Wang, S.~Qiu, A.~Bozoki, and M.~Liu, ``Source-free
  collaborative domain adaptation via multi-perspective feature enrichment for
  functional mri analysis,'' \emph{Pattern Recognition}, vol. 157, p. 110912,
  2025. [Online]. Available:
  \url{https://www.sciencedirect.com/science/article/pii/S0031320324006630}
\BIBentrySTDinterwordspacing

\bibitem{dong2020compression}
D.~Dong, C.~Luo, X.~Guell, Y.~Wang, H.~He, M.~Duan, S.~B. Eickhoff, and D.~Yao,
  ``Compression of cerebellar functional gradients in schizophrenia,''
  \emph{Schizophrenia bulletin}, vol.~46, no.~5, pp. 1282--1295, 2020.

\bibitem{luppi2022synergistic}
A.~I. Luppi, P.~A. Mediano, F.~E. Rosas, N.~Holland, T.~D. Fryer, J.~T.
  O’Brien, J.~B. Rowe, D.~K. Menon, D.~Bor, and E.~A. Stamatakis, ``A
  synergistic core for human brain evolution and cognition,'' \emph{Nature
  Neuroscience}, vol.~25, no.~6, pp. 771--782, 2022.

\bibitem{dong2024brain}
C.~Dong and D.~Sun, ``Brain network classification based on dynamic graph
  attention information bottleneck,'' \emph{Computer Methods and Programs in
  Biomedicine}, vol. 243, p. 107913, 2024.

\bibitem{wang2023covariance}
Y.~Wang, S.~Genon, D.~Dong, F.~Zhou, C.~Li, D.~Yu, K.~Yuan, Q.~He, J.~Qiu,
  T.~Feng \emph{et~al.}, ``Covariance patterns between sleep health domains and
  distributed intrinsic functional connectivity,'' \emph{Nature
  Communications}, vol.~14, no.~1, p. 7133, 2023.

\bibitem{SONG2024BrainDAS}
R.~Song, P.~Cao, G.~Wen, P.~Zhao, Z.~Huang, X.~Zhang, J.~Yang, and O.~R.
  Zaiane, ``Braindas: Structure-aware domain adaptation network for multi-site
  brain network analysis,'' \emph{Medical Image Analysis}, vol.~96, p. 103211,
  2024.

\bibitem{YE2025Fuse}
J.~Ye, Y.~Li, A.~Zeng, and D.~Pan, ``Fuse-former: An interpretability analysis
  model for rs-fmri based on multi-scale information fusion interaction,''
  \emph{Biomedical Signal Processing and Control}, vol. 105, p. 107471, 2025.

\bibitem{Kouw2019Review}
W.~M. Kouw and M.~Loog, ``A review of domain adaptation without target
  labels,'' \emph{IEEE Transactions on Pattern Analysis and Machine
  Intelligence}, vol.~43, no.~3, pp. 766--785, 2021.

\bibitem{Zhao2020aReview}
S.~Zhao, X.~Yue, S.~Zhang, B.~Li, H.~Zhao, B.~Wu, R.~Krishna, J.~E. Gonzalez,
  A.~L. Sangiovanni-Vincentelli, S.~A. Seshia, and K.~Keutzer, ``A review of
  single-source deep unsupervised visual domain adaptation,'' \emph{IEEE
  Transactions on Neural Networks and Learning Systems}, vol.~33, no.~2, pp.
  473--493, 2022.

\bibitem{SUN2015survey}
\BIBentryALTinterwordspacing
S.~Sun, H.~Shi, and Y.~Wu, ``A survey of multi-source domain adaptation,''
  \emph{Information Fusion}, vol.~24, pp. 84--92, 2015. [Online]. Available:
  \url{https://www.sciencedirect.com/science/article/pii/S1566253514001316}
\BIBentrySTDinterwordspacing

\bibitem{chu2022resting}
Y.~Chu, H.~Ren, L.~Qiao, and M.~Liu, ``Resting-state functional mri adaptation
  with attention graph convolution network for brain disorder identification,''
  \emph{Brain Sciences}, vol.~12, no.~10, p. 1413, 2022.

\bibitem{sun2016deep}
B.~Sun and K.~Saenko, ``Deep coral: Correlation alignment for deep domain
  adaptation,'' in \emph{Computer Vision--ECCV 2016 Workshops: Amsterdam, The
  Netherlands, October 8-10 and 15-16, 2016, Proceedings, Part III 14}.\hskip
  1em plus 0.5em minus 0.4em\relax Springer, 2016, pp. 443--450.

\bibitem{FANG2023Unsupervised}
\BIBentryALTinterwordspacing
Y.~Fang, M.~Wang, G.~G. Potter, and M.~Liu, ``Unsupervised cross-domain
  functional mri adaptation for automated major depressive disorder
  identification,'' \emph{Medical Image Analysis}, vol.~84, p. 102707, 2023.
  [Online]. Available:
  \url{https://www.sciencedirect.com/science/article/pii/S1361841522003358}
\BIBentrySTDinterwordspacing

\bibitem{Gretton2006Kernel}
\BIBentryALTinterwordspacing
A.~Gretton, K.~Borgwardt, M.~Rasch, B.~Sch\"{o}lkopf, and A.~Smola, ``A kernel
  method for the two-sample-problem,'' in \emph{Advances in Neural Information
  Processing Systems}, B.~Sch\"{o}lkopf, J.~Platt, and T.~Hoffman, Eds.,
  vol.~19.\hskip 1em plus 0.5em minus 0.4em\relax MIT Press, 2006. [Online].
  Available:
  \url{https://proceedings.neurips.cc/paper_files/paper/2006/file/e9fb2eda3d9c55a0d89c98d6c54b5b3e-Paper.pdf}
\BIBentrySTDinterwordspacing

\bibitem{LI2020Multisite}
\BIBentryALTinterwordspacing
X.~Li, Y.~Gu, N.~Dvornek, L.~H. Staib, P.~Ventola, and J.~S. Duncan,
  ``Multi-site fmri analysis using privacy-preserving federated learning and
  domain adaptation: Abide results,'' \emph{Medical Image Analysis}, vol.~65,
  p. 101765, 2020. [Online]. Available:
  \url{https://www.sciencedirect.com/science/article/pii/S1361841520301298}
\BIBentrySTDinterwordspacing

\bibitem{Ganin2016Domain}
\BIBentryALTinterwordspacing
Y.~Ganin, E.~Ustinova, H.~Ajakan, P.~Germain, H.~Larochelle, F.~Laviolette,
  M.~March, and V.~Lempitsky, ``Domain-adversarial training of neural
  networks,'' \emph{Journal of Machine Learning Research}, vol.~17, no.~59, pp.
  1--35, 2016. [Online]. Available:
  \url{http://jmlr.org/papers/v17/15-239.html}
\BIBentrySTDinterwordspacing

\bibitem{masoudnia2014mixture}
S.~Masoudnia and R.~Ebrahimpour, ``Mixture of experts: a literature survey,''
  \emph{Artificial Intelligence Review}, vol.~42, pp. 275--293, 2014.

\bibitem{Liu2023DomainA}
X.~Liu, J.~Wu, W.~Li, Q.~Liu, L.~Tian, and H.~Huang, ``Domain adaptation via
  low rank and class discriminative representation for autism spectrum disorder
  identification: A multi-site fmri study,'' \emph{IEEE Transactions on Neural
  Systems and Rehabilitation Engineering}, vol.~31, pp. 806--817, 2023.

\bibitem{Baker2024Hyperbolic}
C.~Baker, I.~Suárez-Méndez, G.~Smith, E.~B. Marsh, M.~Funke, J.~C. Mosher,
  F.~Maestú, M.~Xu, and D.~Pantazis, ``Hyperbolic graph embedding of meg brain
  networks to study brain alterations in individuals with subjective cognitive
  decline,'' \emph{IEEE Journal of Biomedical and Health Informatics}, vol.~28,
  no.~12, pp. 7357--7368, 2024.

\bibitem{Dai2021CVPR}
J.~Dai, Y.~Wu, Z.~Gao, and Y.~Jia, ``A hyperbolic-to-hyperbolic graph
  convolutional network,'' in \emph{Proceedings of the IEEE/CVF Conference on
  Computer Vision and Pattern Recognition (CVPR)}, June 2021, pp. 154--163.

\bibitem{yang2024hypformer}
M.~Yang, H.~Verma, D.~C. Zhang, J.~Liu, I.~King, and R.~Ying, ``Hypformer:
  Exploring efficient transformer fully in hyperbolic space,'' in
  \emph{Proceedings of the 30th ACM SIGKDD Conference on Knowledge Discovery
  and Data Mining}, 2024, pp. 3770--3781.

\bibitem{gu2024Unsupervised}
X.~Gu, J.~Sun, and Z.~Xu, ``Unsupervised and semi-supervised robust spherical
  space domain adaptation,'' \emph{IEEE Transactions on Pattern Analysis and
  Machine Intelligence}, vol.~46, no.~3, pp. 1757--1774, 2024.

\bibitem{yang2023Hyperbolic}
M.~Yang, M.~Zhou, H.~Xiong, and I.~King, ``Hyperbolic temporal network
  embedding,'' \emph{IEEE Transactions on Knowledge and Data Engineering},
  vol.~35, no.~11, pp. 11\,489--11\,502, 2023.

\bibitem{Nickel2017Poincar}
\BIBentryALTinterwordspacing
M.~Nickel and D.~Kiela, ``Poincar\'{e} embeddings for learning hierarchical
  representations,'' in \emph{Advances in Neural Information Processing
  Systems}, I.~Guyon, U.~V. Luxburg, S.~Bengio, H.~Wallach, R.~Fergus,
  S.~Vishwanathan, and R.~Garnett, Eds., vol.~30.\hskip 1em plus 0.5em minus
  0.4em\relax Curran Associates, Inc., 2017. [Online]. Available:
  \url{https://proceedings.neurips.cc/paper_files/paper/2017/file/59dfa2df42d9e3d41f5b02bfc32229dd-Paper.pdf}
\BIBentrySTDinterwordspacing

\bibitem{Chami2019Hyperbolic}
I.~Chami, Z.~Ying, C.~R\'{e}, and J.~Leskovec, ``Hyperbolic graph convolutional
  neural networks,'' in \emph{Advances in Neural Information Processing
  Systems}, H.~Wallach, H.~Larochelle, A.~Beygelzimer, F.~d\textquotesingle
  Alch\'{e}-Buc, E.~Fox, and R.~Garnett, Eds.\hskip 1em plus 0.5em minus
  0.4em\relax Curran Associates, Inc., 2019.

\bibitem{Ganea2018Hyperbolic}
O.~Ganea, G.~Becigneul, and T.~Hofmann, ``Hyperbolic neural networks,'' in
  \emph{Advances in Neural Information Processing Systems}, S.~Bengio,
  H.~Wallach, H.~Larochelle, K.~Grauman, N.~Cesa-Bianchi, and R.~Garnett, Eds.,
  vol.~31.\hskip 1em plus 0.5em minus 0.4em\relax Curran Associates, Inc.,
  2018.

\bibitem{guell2018functional}
X.~Guell, J.~D. Schmahmann, J.~D. Gabrieli, and S.~S. Ghosh, ``Functional
  gradients of the cerebellum,'' \emph{elife}, vol.~7, p. e36652, 2018.

\bibitem{huntenburg2018large}
J.~M. Huntenburg, P.-L. Bazin, and D.~S. Margulies, ``Large-scale gradients in
  human cortical organization,'' \emph{Trends in cognitive sciences}, vol.~22,
  no.~1, pp. 21--31, 2018.

\bibitem{guo2023functional}
S.~Guo, L.~Feng, R.~Ding, S.~Long, H.~Yang, X.~Gong, J.~Lu, and D.~Yao,
  ``Functional gradients in prefrontal regions and somatomotor networks reflect
  the effect of music training experience on cognitive aging,'' \emph{Cerebral
  Cortex}, vol.~33, no.~12, pp. 7506--7517, 2023.

\bibitem{gong2023connectivity}
Z.-Q. Gong and X.-N. Zuo, ``Connectivity gradients in spontaneous brain
  activity at multiple frequency bands,'' \emph{Cerebral Cortex}, vol.~33,
  no.~17, pp. 9718--9728, 2023.

\bibitem{hong2019atypical}
S.-J. Hong, R.~Vos~de Wael, R.~A. Bethlehem, S.~Lariviere, C.~Paquola, S.~L.
  Valk, M.~P. Milham, A.~Di~Martino, D.~S. Margulies, J.~Smallwood
  \emph{et~al.}, ``Atypical functional connectome hierarchy in autism,''
  \emph{Nature communications}, vol.~10, no.~1, p. 1022, 2019.

\bibitem{urchs2022functional}
S.~G. Urchs, A.~Tam, P.~Orban, C.~Moreau, Y.~Benhajali, H.~D. Nguyen, A.~C.
  Evans, and P.~Bellec, ``Functional connectivity subtypes associate robustly
  with asd diagnosis,'' \emph{Elife}, vol.~11, p. e56257, 2022.

\bibitem{di2014autism}
A.~Di~Martino, C.-G. Yan, Q.~Li, E.~Denio, F.~X. Castellanos, K.~Alaerts, J.~S.
  Anderson, M.~Assaf, S.~Y. Bookheimer, M.~Dapretto \emph{et~al.}, ``The autism
  brain imaging data exchange: towards a large-scale evaluation of the
  intrinsic brain architecture in autism,'' \emph{Molecular psychiatry},
  vol.~19, no.~6, pp. 659--667, 2014.

\bibitem{luo2024hierarchical}
Y.~Luo, Q.~Chen, F.~Li, L.~Yi, P.~Xu, and Y.~Zhang, ``Hierarchical feature
  extraction on functional brain networks for autism spectrum disorder
  identification with resting-state fmri data,'' \emph{arXiv preprint
  arXiv:2412.02424}, 2024.

\bibitem{vos2020brainspace}
R.~Vos~de Wael, O.~Benkarim, C.~Paquola, S.~Lariviere, J.~Royer, S.~Tavakol,
  T.~Xu, S.-J. Hong, G.~Langs, S.~Valk \emph{et~al.}, ``Brainspace: a toolbox
  for the analysis of macroscale gradients in neuroimaging and connectomics
  datasets,'' \emph{Communications biology}, vol.~3, no.~1, p. 103, 2020.

\bibitem{yan2010dparsf}
C.~Yan and Y.~Zang, ``Dparsf: a matlab toolbox for" pipeline" data analysis of
  resting-state fmri,'' \emph{Frontiers in systems neuroscience}, vol.~4, p.
  1377, 2010.

\bibitem{tzourio2002automated}
N.~Tzourio-Mazoyer, B.~Landeau, D.~Papathanassiou, F.~Crivello, O.~Etard,
  N.~Delcroix, B.~Mazoyer, and M.~Joliot, ``Automated anatomical labeling of
  activations in spm using a macroscopic anatomical parcellation of the mni mri
  single-subject brain,'' \emph{Neuroimage}, vol.~15, no.~1, pp. 273--289,
  2002.

\bibitem{rodriguez2020hyperbolic}
M.~A. Rodr{\'\i}guez-Flores and F.~Papadopoulos, ``Hyperbolic mapping of human
  proximity networks,'' \emph{Scientific reports}, vol.~10, no.~1, p. 20244,
  2020.

\bibitem{Tong2023fMRI}
W.~Tong, Y.-X. Li, X.-Y. Zhao, Q.-Q. Chen, Y.-B. Gao, P.~Li, and E.~Q. Wu,
  ``fmri-based brain disease diagnosis: A graph network approach,'' \emph{IEEE
  Transactions on Medical Robotics and Bionics}, vol.~5, no.~2, pp. 312--322,
  2023.

\bibitem{LIAO2017Individual}
\BIBentryALTinterwordspacing
X.~Liao, M.~Cao, M.~Xia, and Y.~He, ``Individual differences and time-varying
  features of modular brain architecture,'' \emph{NeuroImage}, vol. 152, pp.
  94--107, 2017. [Online]. Available:
  \url{https://www.sciencedirect.com/science/article/pii/S1053811917301775}
\BIBentrySTDinterwordspacing

\bibitem{sporns2016modular}
O.~Sporns and R.~F. Betzel, ``Modular brain networks,'' \emph{Annual review of
  psychology}, vol.~67, no.~1, pp. 613--640, 2016.

\bibitem{van2009functionally}
M.~P. Van Den~Heuvel, R.~C. Mandl, R.~S. Kahn, and H.~E. Hulshoff~Pol,
  ``Functionally linked resting-state networks reflect the underlying
  structural connectivity architecture of the human brain,'' \emph{Human brain
  mapping}, vol.~30, no.~10, pp. 3127--3141, 2009.

\bibitem{van2010exploring}
M.~P. Van Den~Heuvel and H.~E.~H. Pol, ``Exploring the brain network: a review
  on resting-state fmri functional connectivity,'' \emph{European
  neuropsychopharmacology}, vol.~20, no.~8, pp. 519--534, 2010.

\bibitem{geerligs2015state}
L.~Geerligs, M.~Rubinov, R.~N. Henson \emph{et~al.}, ``State and trait
  components of functional connectivity: individual differences vary with
  mental state,'' \emph{Journal of Neuroscience}, vol.~35, no.~41, pp.
  13\,949--13\,961, 2015.

\bibitem{canario2021review}
E.~Canario, D.~Chen, and B.~Biswal, ``A review of resting-state fmri and its
  use to examine psychiatric disorders,'' \emph{Psychoradiology}, vol.~1,
  no.~1, pp. 42--53, 2021.

\bibitem{gallen2019brain}
C.~L. Gallen and M.~D’Esposito, ``Brain modularity: a biomarker of
  intervention-related plasticity,'' \emph{Trends in cognitive sciences},
  vol.~23, no.~4, pp. 293--304, 2019.

\bibitem{Wang2024Leveraging}
Q.~Wang, W.~Wang, Y.~Fang, P.-T. Yap, H.~Zhu, H.-J. Li, L.~Qiao, and M.~Liu,
  ``Leveraging brain modularity prior for interpretable representation learning
  of fmri,'' \emph{IEEE Transactions on Biomedical Engineering}, vol.~71,
  no.~8, pp. 2391--2401, 2024.

\bibitem{Ding2024LGGNet}
Y.~Ding, N.~Robinson, C.~Tong, Q.~Zeng, and C.~Guan, ``Lggnet: Learning from
  local-global-graph representations for brain–computer interface,''
  \emph{IEEE Transactions on Neural Networks and Learning Systems}, vol.~35,
  no.~7, pp. 9773--9786, 2024.

\bibitem{Yan2020Weighted}
H.~Yan, Z.~Li, Q.~Wang, P.~Li, Y.~Xu, and W.~Zuo, ``Weighted and class-specific
  maximum mean discrepancy for unsupervised domain adaptation,'' \emph{IEEE
  Transactions on Multimedia}, vol.~22, no.~9, pp. 2420--2433, 2020.

\bibitem{TANG2022Multitarget}
\BIBentryALTinterwordspacing
L.~Tang, Q.~Zhang, J.~Xuan, T.~Shi, and R.~Li, ``Multitarget domain adaptation
  with transferable hyperbolic prototypes for intelligent fault diagnosis,''
  \emph{Knowledge-Based Systems}, vol. 257, p. 109952, 2022. [Online].
  Available:
  \url{https://www.sciencedirect.com/science/article/pii/S0950705122010450}
\BIBentrySTDinterwordspacing

\bibitem{zhang2020transport}
J.~Zhang, P.~Wan, and D.~Zhang, ``Transport-based joint distribution alignment
  for multi-site autism spectrum disorder diagnosis using resting-state fmri,''
  in \emph{Medical Image Computing and Computer Assisted Intervention--MICCAI
  2020: 23rd International Conference, Lima, Peru, October 4--8, 2020,
  Proceedings, Part II 23}.\hskip 1em plus 0.5em minus 0.4em\relax Springer,
  2020, pp. 444--453.

\bibitem{Wang2020Identifying}
M.~Wang, D.~Zhang, J.~Huang, P.-T. Yap, D.~Shen, and M.~Liu, ``Identifying
  autism spectrum disorder with multi-site fmri via low-rank domain
  adaptation,'' \emph{IEEE Transactions on Medical Imaging}, vol.~39, no.~3,
  pp. 644--655, 2020.

\bibitem{Kunda2023Improving}
M.~Kunda, S.~Zhou, G.~Gong, and H.~Lu, ``Improving multi-site autism
  classification via site-dependence minimization and second-order functional
  connectivity,'' \emph{IEEE Transactions on Medical Imaging}, vol.~42, no.~1,
  pp. 55--65, 2023.

\bibitem{jiang2020hi}
H.~Jiang, P.~Cao, M.~Xu, J.~Yang, and O.~Zaiane, ``Hi-gcn: A hierarchical graph
  convolution network for graph embedding learning of brain network and brain
  disorders prediction,'' \emph{Computers in Biology and Medicine}, vol. 127,
  p. 104096, 2020.

\bibitem{kawahara2017brainnetcnn}
J.~Kawahara, C.~J. Brown, S.~P. Miller, B.~G. Booth, V.~Chau, R.~E. Grunau,
  J.~G. Zwicker, and G.~Hamarneh, ``Brainnetcnn: Convolutional neural networks
  for brain networks; towards predicting neurodevelopment,'' \emph{NeuroImage},
  vol. 146, pp. 1038--1049, 2017.

\bibitem{wen2022mvs}
G.~Wen, P.~Cao, H.~Bao, W.~Yang, T.~Zheng, and O.~Zaiane, ``Mvs-gcn: A prior
  brain structure learning-guided multi-view graph convolution network for
  autism spectrum disorder diagnosis,'' \emph{Computers in biology and
  medicine}, vol. 142, p. 105239, 2022.

\bibitem{luo2023aided}
Y.~Luo, N.~Li, Y.~Pan, W.~Qiu, L.~Xiong, and Y.~Zhang, ``Aided diagnosis of
  autism spectrum disorder based on a mixed neural network model,'' in
  \emph{International Conference on Neural Information Processing}.\hskip 1em
  plus 0.5em minus 0.4em\relax Springer, 2023, pp. 150--161.

\bibitem{Lee2013Resting}
\BIBentryALTinterwordspacing
M.~Lee, C.~Smyser, and J.~Shimony, ``Resting-state fmri: A review of methods
  and clinical applications,'' \emph{American Journal of Neuroradiology},
  vol.~34, no.~10, pp. 1866--1872, 2013. [Online]. Available:
  \url{http://www.ajnr.org/content/34/10/1866}
\BIBentrySTDinterwordspacing

\bibitem{chein2005neuroimaging}
J.~M. Chein and W.~Schneider, ``Neuroimaging studies of practice-related
  change: fmri and meta-analytic evidence of a domain-general control network
  for learning,'' \emph{Cognitive Brain Research}, vol.~25, no.~3, pp.
  607--623, 2005.

\bibitem{raichle2001default}
M.~E. Raichle, A.~M. MacLeod, A.~Z. Snyder, W.~J. Powers, D.~A. Gusnard, and
  G.~L. Shulman, ``A default mode of brain function,'' \emph{Proceedings of the
  national academy of sciences}, vol.~98, no.~2, pp. 676--682, 2001.

\bibitem{corbetta2002control}
M.~Corbetta and G.~L. Shulman, ``Control of goal-directed and stimulus-driven
  attention in the brain,'' \emph{Nature reviews neuroscience}, vol.~3, no.~3,
  pp. 201--215, 2002.

\bibitem{mosconi2015role}
M.~W. Mosconi, Z.~Wang, L.~M. Schmitt, P.~Tsai, and J.~A. Sweeney, ``The role
  of cerebellar circuitry alterations in the pathophysiology of autism spectrum
  disorders,'' \emph{Frontiers in neuroscience}, vol.~9, p. 156522, 2015.

\bibitem{hampson2015autism}
D.~R. Hampson and G.~J. Blatt, ``Autism spectrum disorders and neuropathology
  of the cerebellum,'' \emph{Frontiers in neuroscience}, vol.~9, p. 420, 2015.

\bibitem{xu2021morphological}
X.~Xu, T.~Wang, W.~Li, H.~Li, B.~Xu, M.~Zhang, L.~Yue, P.~Wang, and S.~Xiao,
  ``Morphological, structural, and functional networks highlight the role of
  the cortical-subcortical circuit in individuals with subjective cognitive
  decline,'' \emph{Frontiers in Aging Neuroscience}, vol.~13, p. 688113, 2021.

\bibitem{Wei2023Autistic}
\BIBentryALTinterwordspacing
L.~Wei, B.~Liu, J.~He, M.~Zhang, and Y.~Huang, ``Autistic spectrum disorders
  diagnose with graph neural networks,'' in \emph{Proceedings of the 31st ACM
  International Conference on Multimedia}, ser. MM '23.\hskip 1em plus 0.5em
  minus 0.4em\relax New York, NY, USA: Association for Computing Machinery,
  2023, p. 8819–8827. [Online]. Available:
  \url{https://doi.org/10.1145/3581783.3613818}
\BIBentrySTDinterwordspacing

\end{thebibliography}

\end{document}